\documentclass[lettersize,journal]{IEEEtran}
\usepackage{amsmath,amsfonts}
\usepackage{algorithm}
\usepackage{algpseudocode}
\usepackage{array}
\usepackage[caption=false,font=normalsize,labelfont=sf,textfont=sf]{subfig}
\usepackage{textcomp}
\usepackage{stfloats}
\usepackage{url}
\usepackage{verbatim}
\usepackage{graphicx}
\usepackage{cite}
\usepackage{xcolor}
\usepackage{multirow}
\usepackage{threeparttable}
\usepackage{booktabs}
\usepackage{colortbl}
\definecolor{lightgray}{gray}{0.9}
\usepackage{amssymb}
\usepackage{wasysym} 
\usepackage{float}
\definecolor{mplGreen}{HTML}{008000}
\definecolor{mplRed}{HTML}{FF0000}

\newcolumntype{L}[1]{>{\raggedright\let\newline\\\arraybackslash\hspace{0pt}}m{#1}}
\newcolumntype{C}[1]{>{\centering\let\newline\\\arraybackslash\hspace{0pt}}m{#1}}
\newcolumntype{R}[1]{>{\raggedleft\let\newline\\\arraybackslash\hspace{0pt}}m{#1}}

\makeatletter
\newlength{\@captionlabelwidth}
\DeclareRobustCommand*{\centeredmultilineincaption}[1]{%
  \settowidth{\linewidth}{%
    \@nameuse{fnum@\@captype}: %
  }%
  \begin{tabular}[t]{@{\hspace{-\@captionlabelwidth}}c@{}}%
    \hspace{\@captionlabelwidth}\ignorespaces
    #1%
  \end{tabular}%
}

\begin{document}

\title{FedTrident: Resilient Road Condition Classification Against Poisoning Attacks in Federated Learning}

\author{
Sheng Liu,~\IEEEmembership{Graduate Student Member,~IEEE} and Panagiotis Papadimitratos,~\IEEEmembership{Fellow,~IEEE}
\thanks{Manuscript received X X, 2026; revised X X, 2026. \textit{(Corresponding author: Sheng Liu.)}}
\thanks{Sheng Liu and Panagiotis Papadimitratos are with the Networked Systems Security Group, KTH Royal Institute of Technology, 114 28 Stockholm, Sweden (e-mail: shengliu@kth.se; papadim@kth.se).}
}

\markboth{IEEE TRANSACTIONS ON X,~Vol.~X, No.~X, X~2026}%
{Shell \MakeLowercase{\textit{et al.}}: A Sample Article Using IEEEtran.cls for IEEE Journals}


\maketitle
 
\begin{abstract}
Federated Learning (FL) has emerged as a transformative paradigm for Intelligent Transportation Systems (ITS). With a server aggregating model parameters from vehicles and distributing an updated model to them, FL can train various models supporting advanced ITS tasks, notably camera-based Road Condition Classification (RCC), in a user-privacy-preserving and edge-resource-efficient manner. However, by enabling collaboration, FL-based RCC exposes the system to adversarial participants launching Targeted Label-Flipping Attacks (TLFAs). Malicious clients (vehicles) can relabel their local training data (e.g., from an actual ``uneven" road to a wrong ``smooth" road), consequently compromising global model predictions and jeopardizing transportation safety. Existing countermeasures against such poisoning attacks fail to maintain resilient model performance near the necessary attack-free levels in various attack scenarios due to: 1) not tailoring poisoned local model detection to TLFAs, 2) not excluding malicious vehicular clients based on historical behavior, and 3) not remedying the already-corrupted global model after exclusion. To close this research gap, we propose FedTrident, which introduces: 1) neuron-wise analysis for local model misbehavior detection (notably including attack goal identification, critical feature extraction, and Gaussian Mixture Model (GMM)-based model clustering and filtering); 2) adaptive client rating for client exclusion according to the local model detection results in each FL round; and 3) machine unlearning for corrupted global model remediation once malicious clients are excluded during FL. Extensive evaluation across diverse FL-RCC models, tasks, and configurations demonstrates that FedTrident can effectively thwart TLFAs, achieving performance comparable to that in attack-free scenarios and outperforming eight baseline countermeasures by 9.49\% and 4.47\% for the two most critical metrics. Moreover, FedTrident is resilient to various malicious client rates, data heterogeneity levels, complicated multi-task, and dynamic attacks.  
\end{abstract}


\section{Introduction}\label{sec_intro}

Intelligent Transportation Systems (ITS) hinge on timely and reliable perception of the road environment, where the automated identification of surface conditions (such as unevenness level, friction magnitude, and material type) plays a crucial role~\cite{CHEN2025125978}. In practice, an autonomous vehicle that anticipates a waterlogged stretch via on-board cameras and a deep neural network (DNN)–based \textit{Road Condition Classification (RCC)} module, can proactively modulate speed, reconfigure traction control, and adjust suspension settings. \textit{Image data} is particularly suitable for RCC~\cite{8569396}, as cameras are widely deployed and cost-effective, yet they can capture fine-grained contextual features that improve classification accuracy.

Training RCC models solely on images collected and stored locally at each vehicle separately would suffer from a skewed perception (e.g., one weather and road type mostly encountered). Shifting to \textit{centralized} data collection and model training is also unsustainable in the long run for three reasons: 1) evolving regulations about user privacy\footnote{https://geospy.ai/}, e.g., GDPR\footnote{https://gdpr-info.eu/} in Europe, CCPA\footnote{https://www.oag.ca.gov/privacy/ccpa} in the United States, and PIPL\footnote{http://en.npc.gov.cn.cdurl.cn/2021-12/29/c\_694559.htm} in China, restrict individual data aggregation and transmission; 2) computation and storage burden is concentrated at the data center (server), leaving on-board resources underutilized; and 3) transmitting raw images implies high bandwidth usage and increased response latency for vehicles. \textit{Federated Learning (FL)}~\cite{mcmahan2017communication, AFM3D, you2023federated}, resolves the tension between regulation considerations and resource utilization. Through iterative client-server exchange on model parameters in cross-device horizontal FL, a high-performance global RCC model can be learned without direct image data sharing~\cite{10606293}, while capturing road conditions across a large-scale deployment with varying environments.

\begin{figure}[t]
\centerline{\includegraphics[width=0.48\textwidth]{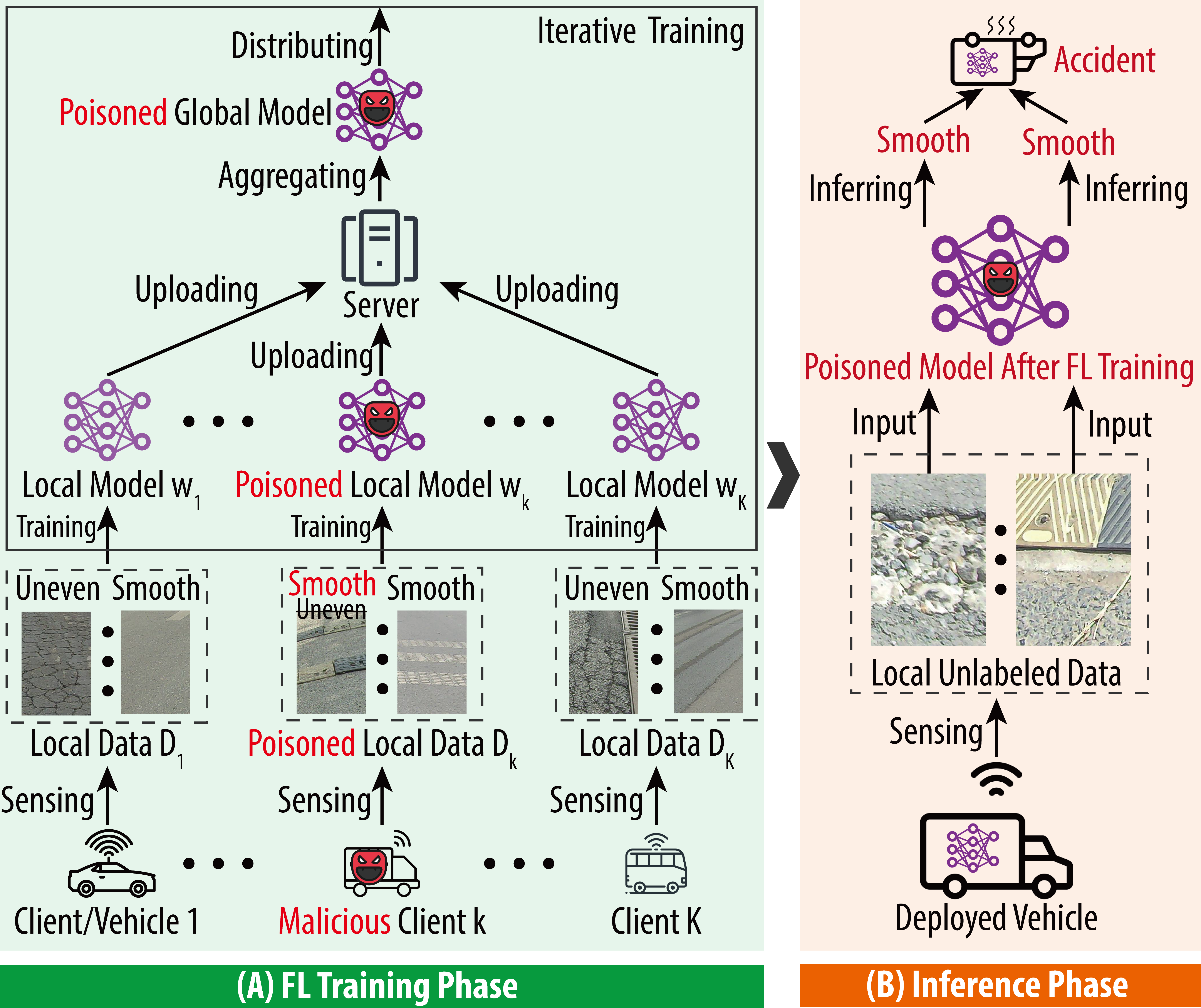}}
\caption{Illustration of TLFAs in FL-RCC systems. (A) Training Phase: Malicious clients deliberately mislabel their data, e.g., from uneven to smooth; thus, their local models are poisoned after local training, and the global model is also poisoned after global aggregation. (B) Inference Phase: Vehicles equipped with the learned model would predict wrong road conditions that threaten transportation safety, e.g., consider actual uneven roads as smooth.}
\label{fig_TLFA_RCC}
\end{figure}

However, FL-based RCC systems, which accept contributions from potentially any participant, are vulnerable to compromised or adversarial clients. \textit{Targeted Label-Flipping Attacks (TLFAs)}~\cite{sameera2024lfgurad} are a particularly potent threat: malicious clients deliberately poisoning their local data by flipping labels from a source class (true) to a target class (falsified). Then, such poisoned data is used for local model training, and its contributions mislead the prediction results of the aggregated global model. For example, as illustrated in Fig.~\ref{fig_TLFA_RCC}, if an adversary relabels road evenness labels from ``uneven” to ``smooth” during the FL training phase, its local model would learn incorrectly from this mislabeled data. Through aggregation rounds in which such an adversary participates, this local corruption progressively poisons the global model. During the inference phase, the consequent RCC model may misclassify actual ``uneven” conditions as ``smooth” in this example. Such an underestimation of hazardous road conditions would increase accident rates and jeopardize transportation safety. Indeed, both findings, e.g., in~\cite{LiuP:C:2025c} and our analysis in Section~\ref{sec_result_anal} demonstrate a catastrophic performance degradation in FL-RCC systems due to TLFAs in the absence of defense mechanisms. TLFAs pose a highly practical threat: unlike more complex attacks (e.g., backdoor attacks~\cite{li2024backdoorindicator}) that require manipulating image pixels or features, TLFAs only necessitate simple label replacement operations, making them easy for adversaries to launch using on-vehicle resources.

Although there are various \textit{countermeasures against poisoning attacks}, they struggle to restore the mis-classification level to that of an attack-free scenario while ensuring robustness in various attack scenarios (even for the state-of-the-art countermeasure against TLFAs for FL-RCC, DEFEND~\cite{LiuP:C:2025d}), because of the following three critical limitations: 

\begin{enumerate}
    \item \textbf{Lack of poisoned local model detection tailored to TLFAs}. Popular defensive mechanisms, such as FoolsGold~\cite{fung2020limitations} and FLAME~\cite{nguyen2022flame}, consider general poisoning attacks; thus, their feature extraction and detection algorithms are not specific to TLFAs. The image data used in FL-RCC are essentially Non-Independent and Identically Distributed (Non-IID), as vehicle driving behaviors vary across locations and over time. General detection methods can easily misidentify poisoned models as benign in such a heterogeneous environment, because legitimate updates themselves also exhibit large variance. Recent poisoning attack mitigation methods primarily focus on backdoor attacks~\cite{fereidooni2023freqfed} or untargeted attacks~\cite{10.1145/3488932.3517395}, rather than TLFAs. FLARE~\cite{LiuP:C:2025c} and DEFEND~\cite{LiuP:C:2025d} only utilize magnitudes for detection; more effective neuron-wise analysis is missing.
    \item \textbf{Shortage of malicious vehicular client exclusion after detection}. Current countermeasures focus on model-level misbehavior detection but overlook vehicle-level exclusion of malicious clients based on those results. When local models are detected and filtered only during training, malicious clients would consistently contribute poisoned models to aggregation, thereby increasing the risk of poisoning and the burden on defensive mechanisms. FLARE~\cite{LiuP:C:2025c} incorporates with a simple count-based client filtering method, and it is moderately effective according to evaluation results. DEFEND~\cite{LiuP:C:2025d} designs partially adaptive client rating with fixed unit rating values; more adaptive exclusion should be developed.
    \item \textbf{Absence of already-corrupted global model remediation beyond detection}. Even if a malicious client is eventually identified and successfully excluded, the state-of-the-art, including DEFEND~\cite{LiuP:C:2025d}, is passive, preventing only current and future poisonous contributions. The malicious updates submitted by the adversary in previous aggregation rounds may already have been deemed benign and integrated into the global model, leaving it in a corrupted state. This ``residual” poison persists within the global model's parameters, continuing to degrade performance and skew predictions long after the attacker is removed. Consequently, any strategy focusing solely on model detection and client exclusion is insufficient~\cite{wu2024unlearning}. An active remediation mechanism is essential to cleanse the global model from past poisoning. The global model recovery literature, e.g., PeriRecover~\cite{10962550}, presumes detection and exclusion, and focuses on remediation. How to integrate remediation together with detection and exclusion is missing.
\end{enumerate}

To fill this gap, building on our earlier work~\cite{LiuP:C:2025c,LiuP:C:2025d}, this paper proposes \textit{FedTrident}, the first defense of its kind that systematically integrates poisoned local model detection, malicious vehicular client exclusion, and corrupted global model remediation for FL-RCC. In each FL round, FedTrident first pinpoints source and target classes (neurons) implicated by TLFAs via neuron-wise analysis of the output layer of each local model. With the two neurons recognized: 1) local model parameters directly connected to them are extracted and then clustered with a Gaussian Mixture Model (GMM) to detect poisoned models, which are then filtered out before global aggregation; 2) two core metrics evaluating TLFAs, Source Recall (SRE)\footnote{The fraction of source class samples that are correctly classified.} and Attack Success Rate (ASR)\footnote{The ratio of samples with source label misclassified into target class.}, are calculated to track the global model performance during training: if performance drops below a threshold, the current global model is still considered contaminated even after the local model filter and thus discarded. Moreover, leveraging the aforementioned poisoned local model detection, FedTrident introduces an adaptive client rating strategy based on decision theory~\cite{PapadimitratosH:J:2006}. A client rating is decremented if its model detection result is bad in a round and incremented otherwise. Once its rating falls below a threshold, the client is identified as malicious with high confidence and promptly excluded from subsequent FL rounds. After exclusion, the adversary's historical contributions to the global aggregation are also removed by utilizing machine unlearning, remedying the corrupted global model. 

Extensive experiments across six DNNs (ResNet-18~\cite{he2016deep}, ResNet-34~\cite{he2016deep}, MobileNet-V3~\cite{howard2019searching}, EfficientNet-B1~\cite{tan2019efficientnet}, DenseNet-121~\cite{huang2017densely}, and Deit-Tiny~\cite{touvron2021training}) and three RCC tasks (friction, material, unevenness classification, and their combination) show that the proposed FedTrident outperforms eight baselines (FedAvg~\cite{mcmahan2017communication}, Krum~\cite{blanchard2017machine}, Trimmed Mean (TMean)~\cite{yin2018byzantine}, Median~\cite{yin2018byzantine}, FoolsGold~\cite{fung2020limitations}, FLAME~\cite{nguyen2022flame}, FLARE~\cite{LiuP:C:2025c}, and DEFEND~\cite{LiuP:C:2025d}) in terms of four evaluation metrics (Global Accuracy (GAC), SRE, ASR, and Global Accuracy of Safety (GAS)). Moreover, FedTrident is resilient to various malicious client rates, diverse data heterogeneity levels, complicated multi-tasks, and dynamic attacks.

In brief, the \textit{main contributions} of this paper are:
\begin{enumerate}
    \item The first complete defense of its kind that systematically integrates poisoned local model detection, malicious vehicular client exclusion, and corrupted global model remediation for FL-RCC against TLFAs.
    \item An effective neuron-wise analysis method for detection, an adaptive client rating method for exclusion, and a retroactive machine unlearning method for remediation.
    \item An extensive evaluation that demonstrates the superiority and resilience of the proposed methods in defending against TLFAs. Notably, when FL-RCC is hit by TLFAs, FedTrident maintains the same model performance as in the attack-free scenario and outperforms the best baseline defense scheme on average by 9.49\% and 4.47\% for SRE and ASR, respectively.
\end{enumerate}

Section~\ref{sec_rw} reviews recent studies on camera-based FL-RCC, TLFAs, and corresponding countermeasures. Section~\ref{sec_sm} details the system and adversary models. Our scheme, FedTrident, is introduced in Section~\ref{sec_DEF} and evaluated in Section~\ref{sec_eva}. Section~\ref{sec_con} concludes and discusses future research. For readability, the key abbreviations used in the paper are listed in TABLE~\ref{tab_abb}.

\section{Background and Related Work}\label{sec_rw}

This section first provides a basic background on FL-RCC, then introduces poisoning attacks that threaten current systems, and countermeasures from three perspectives: detection, exclusion, and remediation. TABLE~\ref{tab_study_sum} summarizes and compares existing solutions. 

\subsection{Federated Learning (FL)-based Road Condition Classification (RCC)}

RCC is a critical function for smart vehicles~\cite{MALIN2019181}. It enables timely adjustments to braking, steering, suspension, and other driver-assistance systems in response to local, unpredictable environmental changes, such as weather conditions. While here we focus on \textit{camera-based RCC}~\cite{zhao2023comprehensive, 10705359}, the task can also rely on other sensors, such as inertial units~\cite{8896015,varona2020deep}. Nonetheless, cameras are or will soon be practically ubiquitous, and equally important, can support proactive detection: a vehicle can perceive and classify conditions such as a water puddle or cracked asphalt well before it makes contact. This contrasts sharply with sensors such as inertial units, which can only supply post-facto data (i.e., after the road puddle or crack is met).

To leverage privacy-sensitive data while utilizing distributed on-board resources, the feasibility of \textit{FL-based RCC} systems was recently explored. FedRD~\cite{YUAN2021385} demonstrates high-performing and privacy-preserving hazardous road damage detection. Follow-up works, FLRSC~\cite{10422129} and FedRSC~\cite{10606293}, further support multi-label RCC tasks within an edge-cloud FL paradigm. However, current FL-RCC proposals primarily focus on improving classification performance and privacy guarantees. Securing these systems against active adversaries remains a largely open problem.

\begin{table}[t]
\centering
\caption{The list of key abbreviations used in this paper.}
\label{tab_abb}
\centering
\renewcommand{\arraystretch}{1.2} 
\begin{tabular}{p{0.131\textwidth}p{0.305\textwidth}}
\hline
\rowcolor{lightgray}
\textbf{Abbreviation}         &\textbf{Meaning}           \\\hline
ITS                    & Intelligent Transportation Systems             \\                  
DNN                    & Deep Neural Network             \\
RCC                    & Road Condition Classification             \\
FL                     & Federated Learning             \\
TLFA                   & Targeted Label-Flipping Attack             \\
Non-IID                & Non-Independent and Identically Distributed             \\                  
GMM                    & Gaussian Mixture Model             \\
SRE                    & Source Recall             \\
ASR                    & Attack Success Rate             \\
GAC                    & Global Accuracy             \\
GAS                    & Global Accuracy of Safety   \\
MPA                    & Model Poisoning Attack     \\
DPA                    & Data Poisoning Attack       \\
SVM                    & Support Vector Machine      \\
KD                     & Knowledge Distillation      \\
VPKI                   & Vehicular Public Key Infrastructure \\
V2X                    & Vehicle to Vehicle/Infrastructure \\
RSCD                   & Road Surface Classification Dataset   \\
\hline
\end{tabular}
\end{table} 

\subsection{Poisoning Attacks Against FL-RCC}

Poisoning attacks in FL can be broadly classified as \textit{Model Poisoning Attacks (MPAs)}~\cite{shejwalkar2021manipulating,11108303} and \textit{Data Poisoning Attacks (DPAs)}~\cite{tolpegin2020data,10054157,khuu2024data}. MPAs demand significant adversarial expertise and computational resources to directly manipulate millions of model parameters~\cite{jebreel2024lfighter,sameera2024lfgurad}. DPAs, in contrast, merely require the simpler operation of altering local data labels, making them more efficient to launch by resource-restricted on-board vehicle platforms. Furthermore, as the FL server can only access local model updates, and not client data, for privacy reasons, the act of data poisoning is, inherently, less visible than model poisoning.

Within the DPA category, attacks can be either untargeted or targeted~\cite{sameera2024lfgurad}. Untargeted DPAs aim to degrade the global model overall performance, but the resultant indiscriminate degradation is often conspicuous, making it more easily detected and countered by the FL system before actual model deployment. Targeted DPAs, notably \textit{TLFAs}~\cite{tolpegin2020data,3670434}, pose a more insidious and significant threat in the context of RCC. On the one hand, a targeted attack is easier to disguise as a benign outlier in a heterogeneous environment. On the other hand, the asymmetric risk inherent in the RCC task allows an adversary to focus on safety-critical misclassifications. To revisit our earlier example, an adversary forcing the model to misclassify ``uneven" as ``smooth" directly jeopardizes vehicle safety. In contrast, the opposite misclassification (``smooth" $\rightarrow$ ``uneven") primarily impacts traffic efficiency. Given their stealthiness, practicality, and high harm potential, TLFAs are the focus of our investigation.

\begin{table*}
\scriptsize
\centering
\caption{Comparative evaluation of available solutions \\ ({\large \CIRCLE} Supported\quad {\large \LEFTcircle} Partially Supported\quad {\large \Circle} Not Supported)}
\label{tab_study_sum}
\renewcommand{\arraystretch}{1.12}
\begin{tabular}{C{2.05cm}|C{1.55cm}|C{1.55cm}|C{1.55cm}|L{9.1cm}} 
\hline
\rowcolor{lightgray}
\multicolumn{1}{c|}{
\renewcommand{\arraystretch}{1.0}
\begin{tabular}{@{}c@{}}
\textbf{Solutions}      
\end{tabular}
} 
&
\multicolumn{1}{c|}{
\renewcommand{\arraystretch}{1.0}
\begin{tabular}{@{}c@{}}
\textbf{Detection$^\ast$}            
\end{tabular}
} 
& 
\multicolumn{1}{c|}{
\renewcommand{\arraystretch}{1.0}
\begin{tabular}{@{}c@{}}
\textbf{Exclusion$^\dagger$}            
\end{tabular}
} 
& 
\multicolumn{1}{c|}{
\renewcommand{\arraystretch}{1.0}
\begin{tabular}{@{}c@{}}
\textbf{Remediation$^{\ddagger}$}            
\end{tabular}
} 
& \multicolumn{1}{c}{
\renewcommand{\arraystretch}{1.0}
\begin{tabular}{@{}c@{}}
\textbf{Highlights}\\
\textbf{(+ pros and - cons)}            
\end{tabular}
} \\ \hline

\begin{tabular}[c]{@{}c@{}}\\  TMean/Median~\cite{yin2018byzantine} \\ \\ \end{tabular}&
\begin{tabular}[c]{@{}c@{}}\\ \large \CIRCLE \\ \\ \end{tabular}&
\begin{tabular}[c]{@{}c@{}}\\ \large \Circle \\ \\ \end{tabular}&
\begin{tabular}[c]{@{}c@{}}\\ \large \Circle \\ \\  \end{tabular}&       
\begin{tabular}{@{}l@{}}
Removes extremes or takes the median of client updates to limit outlier influence: \\
\textbf{+} Simple, could work when most clients are honest and under IID data \\ 
\textbf{-} Vulnerable to optimization-based and colluding attacks as well as Non-IID data \\   
\end{tabular}                               
\\  \hline

\begin{tabular}[c]{@{}c@{}} \\ FoolsGold~\cite{fung2020limitations} \\ \\ \end{tabular}&
\begin{tabular}[c]{@{}c@{}} \\ \large \CIRCLE \\ \\ \end{tabular}&  
\begin{tabular}[c]{@{}c@{}} \\ \large \Circle \\ \\ \end{tabular}&
\begin{tabular}[c]{@{}c@{}} \\ \large \Circle \\ \\ \end{tabular}&
\begin{tabular}{@{}l@{}}
Down-weights clients whose gradient histories are overly similar: \\
\textbf{+} Directly exploits behavioral similarity as a robust, model-agnostic signal \\
\textbf{-} Effectiveness drops with sophisticated diversification and non-Sybil strategies \\                     
\end{tabular}                                                    
\\  \hline

\begin{tabular}[c]{@{}c@{}}\\ FLAME~\cite{nguyen2022flame} \\ \\ \end{tabular}&  
\begin{tabular}[c]{@{}c@{}}\\ \large \LEFTcircle \\ \\ \end{tabular}&    
\begin{tabular}[c]{@{}c@{}}\\ \large \Circle \\ \\ \end{tabular}&   
\begin{tabular}[c]{@{}c@{}}\\ \large \Circle \\ \\ \end{tabular}&  
\begin{tabular}{@{}l@{}}
Clusters client updates and injects calibrated noise plus clipping to erase backdoor: \\
\textbf{+} Demonstrates practical backdoor mitigation with modest impact on benign accuracy \\
\textbf{-} Challenged by adaptive attacks and extreme heterogeneity \\         
\end{tabular}                                                         
\\  \hline

\begin{tabular}[c]{@{}c@{}} \\ FreqFed~\cite{fereidooni2023freqfed} \\ \\ \end{tabular}&
\begin{tabular}[c]{@{}c@{}} \\ \large \LEFTcircle \\ \\ \end{tabular}&  
\begin{tabular}[c]{@{}c@{}} \\ \large \Circle \\ \\ \end{tabular}&
\begin{tabular}[c]{@{}c@{}} \\ \large \Circle \\ \\ \end{tabular}&
\begin{tabular}{@{}l@{}}
Uses frequency-domain analysis of model updates to distinguish poisoned contributions: \\
\textbf{+} Effectively exposes backdoor and untargeted attacks  \\
\textbf{-} Heavily tied to assumed spectral patterns and model structures \\                    
\end{tabular}                                                    
\\  \hline

\begin{tabular}[c]{@{}c@{}} \\ LFGurad~\cite{sameera2024lfgurad} \\ \\ \end{tabular}&
\begin{tabular}[c]{@{}c@{}} \\ \large \CIRCLE \\ \\ \end{tabular}&  
\begin{tabular}[c]{@{}c@{}} \\ \large \Circle \\ \\ \end{tabular}&
\begin{tabular}[c]{@{}c@{}} \\ \large \Circle \\ \\ \end{tabular}&
\begin{tabular}{@{}l@{}}
Uses last-layer activations with an SVM to flag malicious participants: \\
\textbf{+} Tailored to vehicular networks with a concrete pipeline for attack detection \\
\textbf{-} Depends on a representative, trusted auxiliary dataset                    
\end{tabular}                                                    
\\  \hline

\begin{tabular}[c]{@{}c@{}} \\ FLDetector~\cite{10.1145/3534678.3539231} \\ \\ \end{tabular}&
\begin{tabular}[c]{@{}c@{}} \\ \large \LEFTcircle \\ \\ \end{tabular}&  
\begin{tabular}[c]{@{}c@{}} \\ \large \LEFTcircle \\ \\ \end{tabular}&
\begin{tabular}[c]{@{}c@{}} \\ \large \Circle \\ \\ \end{tabular}&
\begin{tabular}{@{}l@{}}
Checks temporal consistency of local updates via Hessian-informed trajectory prediction: \\
\textbf{+} Works without a clean validation set on the server side \\
\textbf{-} Fragile under strong non-IID and large number of malicious clients                    
\end{tabular}                                                    
\\  \hline

\begin{tabular}[c]{@{}c@{}}\\ UL~\cite{wu2024unlearning}  \\ \\ \end{tabular}&
\begin{tabular}[c]{@{}c@{}}\\ \large \Circle \\ \\ \end{tabular}&
\begin{tabular}[c]{@{}c@{}}\\ \large \Circle \\ \\ \end{tabular}&
\begin{tabular}[c]{@{}c@{}}\\ \large \LEFTcircle \\ \\ \end{tabular}&  
\begin{tabular}{@{}l@{}}
Removes backdoor by historical update subtraction and improves accuracy by KD: \\
\textbf{+} Requires no client participation and no knowledge of triggers \\
\textbf{-} TLFAs, Non-IID data, and large-scale FL systems are unconsidered \\         
\end{tabular}                                                      
\\  \hline

\begin{tabular}[c]{@{}c@{}}\\ PeriRecover~\cite{10962550} \\ \\ \end{tabular}&
\begin{tabular}[c]{@{}c@{}}\\ \large \Circle \\ \\ \end{tabular}&
\begin{tabular}[c]{@{}c@{}}\\ \large \Circle \\ \\ \end{tabular}&
\begin{tabular}[c]{@{}c@{}}\\ \large \CIRCLE \\ \\ \end{tabular}&  
\begin{tabular}{@{}l@{}}
Periodically reconstructs a clean model by reusing stored training states: \\
\textbf{+} Achieves near-train-from-scratch recovery within short time \\
\textbf{-} Requires extensive logging of training history and rests on strong smooth assumptions \\         
\end{tabular}                                                      
\\  \hline

\begin{tabular}[c]{@{}c@{}}\\ FLARE~\cite{LiuP:C:2025c} \\ \\ \end{tabular}&
\begin{tabular}[c]{@{}c@{}}\\ \large \CIRCLE \\ \\ \end{tabular}&
\begin{tabular}[c]{@{}c@{}}\\ \large \CIRCLE \\ \\ \end{tabular}&
\begin{tabular}[c]{@{}c@{}}\\ \large \Circle \\ \\ \end{tabular}&  
\begin{tabular}{@{}l@{}}
Combines cluster-based misbehavior detection with count-based client filter: \\
\textbf{+} Improves performance compared to other baselines \\
\textbf{-} Still leaves a noticeable gap to the no-attack baseline and remediation is untouched \\         
\end{tabular}                                                      
\\  \hline

\begin{tabular}[c]{@{}c@{}}\\ DEFEND~\cite{LiuP:C:2025d} \\ \\ \end{tabular}&
\begin{tabular}[c]{@{}c@{}}\\ \large \CIRCLE \\ \\ \end{tabular}&
\begin{tabular}[c]{@{}c@{}}\\ \large \CIRCLE \\ \\ \end{tabular}&
\begin{tabular}[c]{@{}c@{}}\\ \large \Circle \\ \\ \end{tabular}&  
\begin{tabular}{@{}l@{}}
Incorporates poisoned model detection with a partial adaptive client filter: \\
\textbf{+} Tailored to TLFAs in FL-RCC, and maintains low ASR performance\\
\textbf{-} Detection misses angle, exclusion lacks adaptation, and remediation is untouched \\         
\end{tabular}                                                      
\\  \hline

\rowcolor{teal!12}
\begin{tabular}[c]{@{}c@{}}\\ \\ FedTrident (Ours)\\ \\ \end{tabular} &
\begin{tabular}[c]{@{}c@{}}\\ \\ \large \CIRCLE \\ \\ \end{tabular} & 
\begin{tabular}[c]{@{}c@{}}\\ \\ \large \CIRCLE \\ \\ \end{tabular} & 
\begin{tabular}[c]{@{}c@{}}\\ \\ \large \CIRCLE \\ \\ \end{tabular} &  
\begin{tabular}{@{}l@{}}
A joint solution integrating detection, exclusion, and remediation for FL-RCC: \\
+ Accurately detects TLFA-poisoned local models  \\ 
+ Adaptively excludes malicious vehicles based on detection results \\
+ Smoothly remediates corrupted global model after each exclusion \\
+ Resilient to various models, malicious client rates, Non-IID levels, and dynamic attacks \\
\end{tabular}    
\\ \hline
\end{tabular}
\begin{tablenotes} 
	\footnotesize
	\item$^\ast$ \textit{Detection} represents poisoned local model detection.
    \item$^\dagger$ \textit{Exclusion} represents malicious vehicular client exclusion. 
    \item$^{\ddagger}$ \textit{Remediation} represents corrupted global model remediation.
\end{tablenotes}
\end{table*}

\subsection{Defensive Mechanisms for FL-RCC}

A variety of defenses against \textit{generic DPAs} in FL have been proposed, yet most overlook the specifics of RCC and TLFAs. Krum~\cite{blanchard2017machine} first computes the sum of Euclidean distances between each local update, then chooses the update with the smallest summed distance as the available one for poisoning attack mitigation. TMean and Median~\cite{yin2018byzantine} utilize statistics (e.g., mean and median) of model parameters to select several representative local models for aggregation, thus filtering out possibly poisoned models and alleviating attack influence. Based on the assumption that poisoned models are more similar to each other than benign models, FoolsGold~\cite{fung2020limitations} calculates the cosine similarity between local model's output layers and then penalizes more similar contributions to mitigate potential negative attacks. CONTRA~\cite{CONTRA} likewise relies on cosine similarity to adjust client credibility. FLAME~\cite{nguyen2022flame} combines differential privacy, model clustering, and weight clipping technologies, improving robustness at the cost of degraded clean accuracy. 

However, FL-RCC inherently operates in a \textit{Non-IID} regime: spatially and temporally diverse driving patterns induce substantial variation in local data distributions. This intrinsic heterogeneity blurs the boundary between benign and malicious behavior~\cite{fereidooni2023freqfed}, with poisoned updates remaining indistinguishable, substantially degrading existing RCC defenses. Therefore, fine-grained feature extraction and TLFA-specific detection explicitly robust to such heterogeneity are required.

Recent novel countermeasures against DPAs have focused on mitigating \textit{backdoor attacks}. Specifically, FreqFed~\cite{fereidooni2023freqfed} capitalizes on the discrete cosine transform to distinguish malicious updates in the frequency domain, while CrowdGuard~\cite{rieger2022crowdguard} analyzes hidden layer outputs and executes iterative pruning. However, these methods are tailored to the specific characteristics of backdoor attacks, wherein adversaries elaborately falsify both the original image data (by adding triggers) and the corresponding labels. As such, they are ill-suited and not directly applicable to our problem, which concerns TLFAs (label-only attacks).

Few defensive mechanisms are designed specifically for \textit{vehicular settings}. LFGurad~\cite{sameera2024lfgurad} introduces a hierarchical FL framework for vehicular networks, classifying malicious models via a multi-class Support Vector Machine (SVM) applied to output-layer activations, evaluated on structured traffic sign classification datasets. RoHFL~\cite{10046398} proposes a robust FL scheme with logarithm-based normalization to mitigate maliciously scaled model parameters. OQFL~\cite{9641742} employs quantum-behaved particle swarm optimization to adapt FL hyperparameters against adversarial attacks in autonomous driving. However, RoHFL and OQFL are primarily evaluated on generic datasets, such as MNIST and Fashion-MNIST, lacking any practical ITS considerations. Moreover, all three methods concentrate on passive, model-level anomaly detection, not considering a proactive mechanism for vehicle-level malicious client exclusion. 

The research on \textit{malicious client exclusion} and \textit{corrupted model remediation} is also limited. By exploiting the temporal consistency of model updates, FLDetector~\cite{10.1145/3534678.3539231} predicts an expected update for each client based on their history. A client is flagged as malicious and deleted if its update consistently deviates from expectations. However, FLDetector does not perform well in realistic Non-IID settings, particularly with very high benign client variance. FLDetector is inefficient because it re-trains the global model from scratch, using the remaining clients after the exclusion, a costly approach. UL~\cite{wu2024unlearning} leverages subtraction of historical updates to remove backdoors from the global model; while using Knowledge Distillation (KD) to improve high clean accuracy. However, it targets backdoor threats, not TLFAs. PeriRecover~\cite{10962550} precomputes diagonals of the Hessian matrix during the original training phase, and later uses this to estimate the ``clean” gradients for periodic recovery. However, PeriRecover requires lots of extra computation and storage costs.

FLARE~\cite{LiuP:C:2025c} is part of the \textit{state-of-the-art} in defending FL-RCC under TLFAs, combining HDBSCAN-based clustering for poisoned model detection with a count-based client filtering rule. However, FLARE provides only moderate protection, leaving a substantial gap between defended and TLFA-free performance. In a no-attack setting, SRE reaches about 80\%, whereas under TLFAs with FLARE deployed, SRE still drops to roughly 60\%; similarly, ASR increases from around 5\% (no attack) to approximately 30\%. By contrast, our FedTrident framework (Section~\ref{sec_result_anal}) attains the two values to the TLFA-free level. DEFEND~\cite{LiuP:C:2025d} incorporates model misbehavior detection with partially adaptive client rating, which is the basis of FedTrident; however, its detection only considers magnitudes while missing angles, and its unit rating values are fixed, not fully adaptive. Moreover, from the perspective of defense integrity, neither FLARE nor DEFEND considers corrupted global model remediation. All in all, under TLFAs, no existing proposal uses joint detection, exclusion, and remediation to maintain resilient performance in various scenarios.

\begin{table}[t]
\centering
\caption{The list of key notations used in this paper.}
\label{tab_nota}
\centering
\renewcommand{\arraystretch}{1.2} 
\begin{tabular}{p{0.09\textwidth}p{0.34\textwidth}}
\hline
\rowcolor{lightgray}
\textbf{Notation}       &\textbf{Meaning}           \\\hline
$K$                    & The total number of clients/vehicles             \\                  
$\mathbb{C}$           & The entire client cluster             \\
$E$                    & the number of road condition classes             \\
$S$                    & The ordered label sequence  \\
$L$                    & The output layer of RCC model             \\
$\mathbb{D}_k$         & The local dataset of client $c_k$              \\
$n_k$                  & The number of samples in dataset $\mathbb{D}_k$             \\                  
$T$                    & The total number of FL rounds             \\
$\mathbb{C}^t$         & The participant sub-cluster in round $t$             \\
$M$                    & The number of participants in a round             \\
$\omega^t$             & The global model in round $t$             \\
$\omega_k^t$           & The updated local model of client $c_k$ in round $t$  \\
$\eta$                 & The learning rate in local training  \\
$q_k$                  & The aggregation weight of client $c_k$  \\
$\mathbb{C}^t_{good}$  & The identified good participant sub-cluster in round $t$  \\
$P$                    & The number of initialized malicious clients  \\
$f^t$                  & The source class in round $t$  \\
$g^t$                  & The target class in round $t$  \\
$d_{i,j}$              & The ``label distance" between true label $s_i$ and predicted label $s_j$ \\
$n_{i,j}$              & The number of test samples with true label $s_i$ predicted as $s_j$ \\
$f^{t'} $              & The identified source class in round $t$  \\
$ g^{t'}$              & The identified target class in round $t$  \\
$\mathbb{C}^t_{bad}$   & The identified bad participant sub-cluster in round $t$   \\
$SRE^{thr}$            & The threshold of SRE for validation  \\
$ASR^{thr}$            & The threshold of ASR for validation  \\
$r^{max}$              & The maximum rating value  \\
$r^{min}$              & The minimum rating value  \\
$\beta$                & The reward value  \\
$\gamma$               & The penalty value \\
$\mathbb{B}$           & The blacklist  \\
\hline
\end{tabular}
\end{table} 

\section{System Model and Adversary Model}\label{sec_sm}

This section describes the system model and adversarial model. For readability, the notations used in this paper are summarized in TABLE~\ref{tab_nota}.

\subsection{System Model}

\textbf{Vehicular Protocols for Secure and Private Communication:} We consider an FL-RCC system comprising a trusted server and a large pool of participating vehicular clients, each enrolled through a cloud-based Vehicular Public Key Infrastructure (VPKI)~\cite{10075082,KhodaeiNP:J:2023}. Vehicles are equipped with short-lived pseudonym certificates~\cite{KhodaeiJP:J:2018}, i.e., syntactically unlinkable credentials with lifetimes of minutes to hours, providing authenticity, integrity, non-repudiation, and privacy (conditional anonymity with long-term unlinkability). This design is aligned with standardized security and privacy requirements for cooperative ITS and V2X Vehicle-to-Vehicle/Infrastructure) communications~\cite{10075082,4689252,KhodaeiP:J:2015}. Misbehavior associated with one or more pseudonyms can be promptly acted upon via pseudonym resolution and revocation mechanisms~\cite{KhodaeiP:J:2021b}. During the FL operation, clients use their currently valid pseudonym to establish TLS channels~\cite{rescorla2018transport} with the server, ensuring the secure and privacy-preserving transmission of model updates.

\textbf{FL Procedure:} Assume a cluster of $K$ clients, $\mathbb{C} = \{c_1, c_2, \ldots, c_K\}$, collaboratively training an RCC model. The task comprises $E$ road condition classes, and they form an ordered sequence $S = [s_1,s_2,\ldots,s_E]$, where larger indices correspond to more hazardous conditions. The output layer, $L$, of the RCC model thus contains $E$ neurons, $L = [l_1, l_2, \ldots, l_E]$. Each client, $c_k$, holds a private dataset $\mathbb{D}_k$ with $n_k$ image–label pairs, where images are captured by on-board cameras and could be labeled via driver feedback or annotation tools \cite{10480248}.

In each FL round $t\leq T$, a subset of clients $\mathbb{C}^t \subseteq \mathbb{C}$ with $|\mathbb{C}^t| = M \leq K$ is randomly selected for participation. After receiving the current global model $\omega^t$ from the server as its current initialized local model $\omega_k^t$, each participant $c_k \in \mathbb{C}^t$ performs local training on $\mathbb{D}_k$ and obtains an updated model $\omega_k^t$ according to Equation (\ref{eq_sgd}),
\begin{equation}\label{eq_sgd}
    \omega_k^t = \omega^t_k - \eta \nabla_{\omega^t_k} \mathcal{L}_k(\omega^t_k, \mathbb{D}_k)
\end{equation}
where $\eta$ is the learning rate and $\mathcal{L}_k$ denotes the local loss function of client $c_k$ (e.g., cross-entropy). In round $t$, Equation (\ref{eq_sgd}) could be executed several times by $c_k$, not just once.

If there are no poisoned local models, the server aggregates all received updates to form the newest global model, $\omega^{t} = \sum_{c_k \in \mathbb{C}^t} q^t_k \, \omega_k^t$, where $q^t_k$ is the aggregation weight of client $c_k$ in round $t$. While FedAvg \cite{mcmahan2017communication} typically sets $q^t_k$ proportional to $n_k$, we adopt a uniform weighting, $q^t_k = \frac{1}{|\mathbb{C}^t_{good}|}$, to prevent adversaries from inflating their influence via falsified data size claims, where $\mathbb{C}^t_{good}$ is the identified good participant sub-cluster in round $t$. If poisoned models are detected, only benign models will be aggregated as $\omega^{t} = \sum_{c_k \in \mathbb{C}^t_{good}} q_k \, \omega_k^t$. This procedure is iterated until convergence, and the resulting global model is deployed to autonomous vehicles to support real-time RCC on unseen road surface images.

\begin{figure*}[tb]
\centerline{\includegraphics[width=0.987\textwidth]{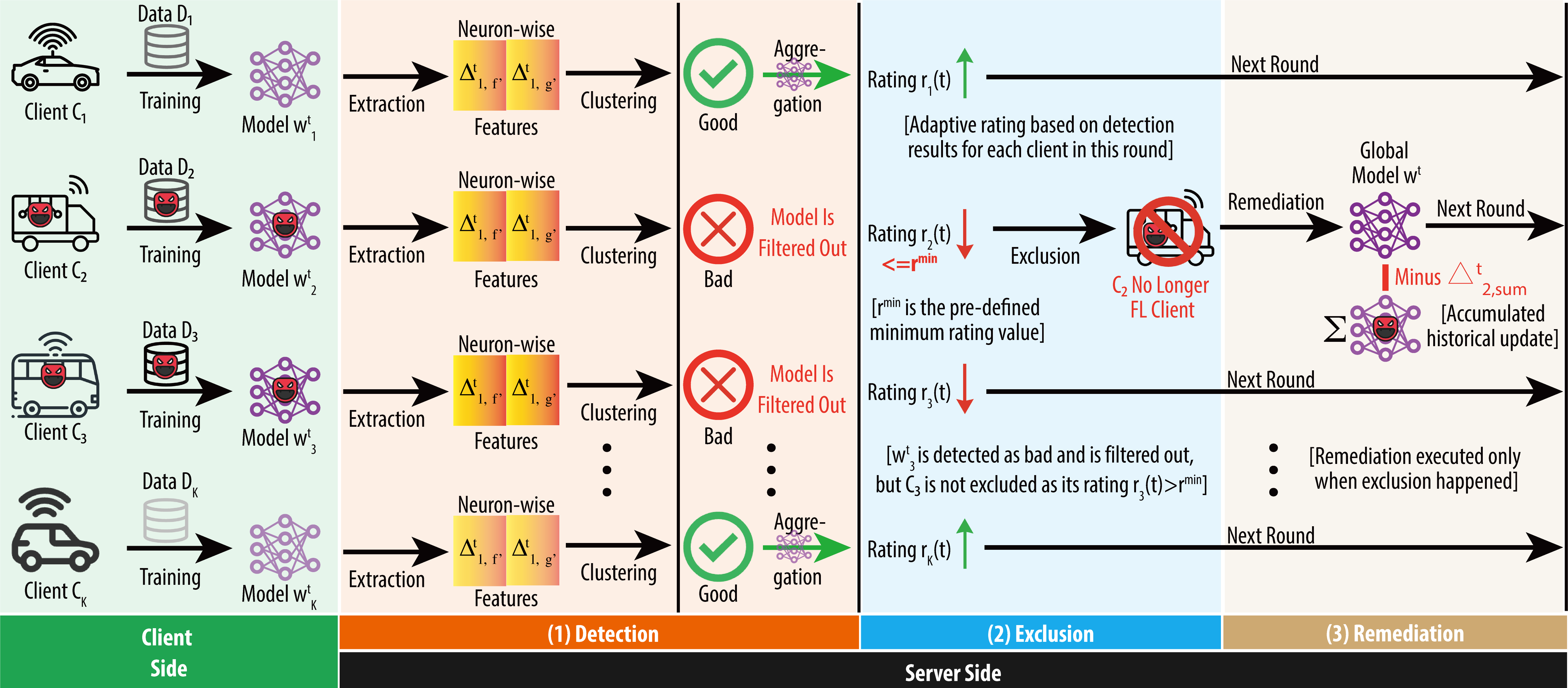}}
\caption{Overview of FedTrident in FL round $t$: (1) poisoned local model detection based on neuron-wise analysis, (2) malicious vehicular client exclusion based on adaptive rating, and (3) corrupted global model remediation based on machine unlearning.}
\label{fig_FedTrident}
\end{figure*}

\subsection{Adversary Model}\label{sec_adv_model}

Assume $P$ malicious clients are present in the system from the outset, with $P < \frac{K}{2}$. Each adversarial client in round $t$ flips labels from a \textit{more hazardous} source class $f^t$ to a \textit{less hazardous} target class $g^t$ (without modifying the input images, and $idx(f^t)>idx(g^t)$), trains its local model on this poisoned dataset, and then submits corrupted updates to the server. Through aggregation, these updates bias the global model so that samples from class $f^t$ are increasingly misclassified as class $g^t$ (e.g., ``uneven" misclassified as ``smooth", as illustrated in Fig.~\ref{fig_TLFA_RCC}), thereby causing vehicles to underestimate road risk. This safety-critical and asymmetric degradation is more harmful than uniform accuracy loss from untargeted poisoning, where no specific hazardous misclassification is enforced. Note that if $idx(f^t)<idx(g^t)$, the attack goal is traffic efficiency rather than safety, e.g., flipping from actual ``smooth" roads to ``uneven". 

Beyond DEFEND~\cite{LiuP:C:2025d}, we assume attackers can adopt dynamic strategies, varying the chosen source and/or target classes across FL rounds, aiming to evade detection and undermine defense mechanisms. For example, considering ``water” images are more similar to ``wet” images than to ``dry” images, attackers can flip from ``water" to ``wet" at the beginning, then from ``water" to ``dry", to bypass defenses and disguise their final target (``dry”). We assume adversaries cannot compromise the trusted server and have no control over the random client selection process, $\mathbb{C}^t$, on the server side. Malicious clients may be introduced gradually as registered entities with valid credentials but with adversarially implemented FL-RCC functionality, or as originally benign clients (vehicles) that are compromised later, e.g., by malware causing them to deviate from the prescribed protocol.

\textbf{Capturing TLFA Impact on Transportation Safety:}
Conventional metrics (e.g., SRE and GAC) treat all misclassifications equally, ignoring the asymmetric impact of TLFAs on safety. For example, misclassifying $s_2$ (wet) or $s_6$ (ice) as $s_1$ (dry) is scored identically by standard metrics, although confusing a highly hazardous class $s_6$ (ice) with a benign class $s_1$ (dry) is substantially more dangerous than confusing $s_2$ (wet) with $s_1$ (dry). Similarly, predicting $s_6$ (ice) as $s_4$ (snow) or as $s_1$ (dry) yields the same penalty, despite the latter representing a much more severe underestimation of transportation safety. This motivates a safety-aware metric that incorporates the \textit{distance} between true and predicted classes in the ordered label space $\mathcal{S}$.

Following the idea in FLARE \cite{LiuP:C:2025c}, we leverage a weighted label distance $d_{i,j}$ between true label $s_i$ and predicted label $s_j$, defined as Equation~(\ref{eq_exp}),

\begin{equation}\label{eq_exp}
    d_{i,j} = \left(\frac{e}{2}\right)^{|i-j|}
\end{equation}
where $e$ is Euler's number. Correct predictions have $d_{i,i} = 1$, while larger index gaps yield exponentially higher penalties, reflecting more severe safety violations.

$d_{i,j}$ is incorporate into a weighted confusion matrix $X$ with element $X[i,j] = d_{i,j} \cdot n_{i,j}$, where $n_{i,j}$ is the number of test samples with true label $s_i$ predicted as $s_j$. Based on $X$, \textit{GAS} is defined as Equation (\ref{eq_error_new}),

\begin{equation}\label{eq_error_new}
    GAS = \frac{\sum_{i=1}^{E} n_{i,i}}{\sum_{i=1}^{E} \sum_{j=1}^{E} d_{i,j} \cdot n_{i,j}}
\end{equation}

Lower GAS values indicate stronger safety degradation, i.e., more dangerous underestimations. The same weighting principle can be applied to derive TLFA-aware variants of SRE, ASR, and other metrics; we adopt GAS as it aggregates the full distance structure across all $(i,j)$ pairs.

\begin{algorithm}[tb]
\caption{Protocol of FedTrident}
\label{alg_1}
\begin{algorithmic}[1]
\State Initialize global model $\omega^0$, blacklist $\mathbb{B} = \emptyset$, $SRE^{old}=0$, $ASR^{old}=1$, and rating value $r_k(0)=\delta(r^{max}-r^{min})$ and $\Delta \omega^{0}_{k,sum}=0$ for each client $c_k$
\For{each round $t \in [1,T]$}
    \State $\mathbb{C}^t \gets$ Randomly select $M$ clients from $\mathbb{C}-\mathbb{B}$
    \State The server sends $\omega^{t-1}$ to all clients in $\mathbb{C}^t$
    \For{each client $c_k \in \mathbb{C}^t$ in parallel}
        \State Update local model $\omega^t_k$
        \State Send $\omega^t_k$ back to the server
    \EndFor
    \State The server receives $\omega^t_k$ from $\mathbb{C}^t$
    \For{each $\omega^t_k$ the server}
        \State $\Delta^t_{k,L}=\{\omega^t_{k,l}-\omega^{t-1}_l|l\in L\}$ \Comment{Parameter changes}
        \State Calculate $\ell_2$-norm magnitudes $||\Delta^t_{k,l}||_2$ for $l\in L$
    \EndFor
    \State $\{||\Delta^t_{l_{1}}||_2,...,||\Delta^t_{l_{E}}||_2\}\gets$ Neuron-wise magnitudes
    \State $\{||S^t_{l_{1}}||_2,...,||S^t_{l_{E}}||_2\}\gets$ Combine angles
    \State $f^{t'}, g^{t'}$ $\gets$ Top-2($\{||S^t_{l_{1}}||_2,...,||S^t_{l_{E}}||_2\}$)\Comment{$f^{t'}<g^{t'}$}
    \State $U^t \gets \{\Delta^t_{k,l}|c_k \in \mathbb{C}^t, l \in \{f^{t'}, g^{t'}\}\}$
    \State $\mathbb{C}^t_{good},\mathbb{C}^t_{bad}$ = GMM($U^t$) \Comment{Detection}
    \State $\omega^{t}=Aggeregate\{\omega^t_{k}|c_k \in \mathbb{C}^t_{good}\}$
    \For{each $\omega^t_k (c_k\in \mathbb{C}^t_{good})$ the server}
        \State $\Delta \omega^t_{k,sum}=\Delta \omega^{t-1}_{k,sum}+(\omega^t_{k}-\omega^{t-1})$
    \EndFor
    \State $SRE^{new}, ASR^{new}$ $\gets$  Validate ($\omega^{t}$)
    \State $\Delta SRE/ASR=SRE/ASR^{new}-SRE/ASR^{old}$
    \If{$\Delta SRE<SRE^{thr}$ or $\Delta ASR>ASR^{thr}$}
        \State $\omega^{t}=\omega^{t-1}$
    \EndIf
    \State $SRE^{old}=SRE^{new}$, $ASR^{old}_k=ASR^{new}_k$
    \For{$c_k \in \mathbb{C}^t$ the server}
        \If{$c_k \in \mathbb{C}^t_{bad}$}
            \State $r_k(t)=max\{r_k(t-1)-\gamma_k^t, r_{min}\}$
            \If{$r_k(t)\leq r^{min}$ $And$ $c_k \notin \mathbb{B}$ }
                \State Add $c_k$ in $\mathbb{B}$ \Comment{Exclusion}
                \State $\omega^{t}=\omega^{t}-\frac{1}{M} \Delta \omega^t_{k,sum}$ \Comment{Remediation}
            \EndIf
        \Else
            \State $r_k(t)=min\{r_k(t-1)+\beta_k^t, r^{max}\}$
        \EndIf
    \EndFor
\EndFor
\State \Return $\omega^{T}$
\end{algorithmic}
\end{algorithm}

\section{FedTrident: Our Scheme}\label{sec_DEF}

In this section, we introduce our proposed FedTrident: a poisoned local model detection, malicious vehicular client exclusion, and corrupted global model remediation mechanism for FL-RCC. Fig.~\ref{fig_FedTrident} illustrates the overall workflow of FedTrident in each round.

\subsection{Scheme Overview} 

The overall protocol is described in Algorithm \ref{alg_1}. \textit{Line 1} initializes a client blacklist $\mathbb{B}$, an old SRE value, an old ASR value, and for each client $c_k$ a rating value $r_k(0)$. \textit{Lines 2-8} randomly select clients not in the blacklist to execute local model training and upload it for each round, protected by security and privacy protocols described in the system model, such as pseudonym certificates and authenticated TLS. \textit{Lines 9-16} analyze neuron-wise magnitudes and angles regarding output layer $L$ to identify source and target classes as $f^{t'}$ and $g^{t'}$. \textit{Lines 17-18} detect poisoned models via a Gaussian Mixture Model (GMM) based on $U^t$: value changes of parameters connected to $f^{t'}$ and $g^{t'}$. \textit{Line 19} forms a new global model $\omega^{t}$ by only aggregating good models. \textit{Lines 20-22} update the accumulated historical contribution $\Delta \omega^t_{k,sum}$ for each good client. \textit{Lines 23-28} validate the new global model $\omega^{t}$ based on SRE and ASR values to decide accept or discard it. \textit{Lines 29-39} update rating values $r_k(t)$ and the blacklist $\mathbb{B}$ to exclude malicious clients. Each of the detected outliers in $\mathbb{C}^t_{out}$, sends model parameters over each of the secure channels in a non-repudiable manner. Given the use of a valid pseudonym (contributions of a given client are anonymized, yet they can be linked to each other across FL rounds), client rating can be reduced, so that a deemed malicious client can be excluded from future FL processes, while rendering a client's participation in different FL executions unlinkable. \textit{Line 34} remediate the corrupted global model after exclusion.

\subsection{Poisoned Local Model Detection} 

\begin{figure}[tb]
\centerline{\includegraphics[width=0.48\textwidth]{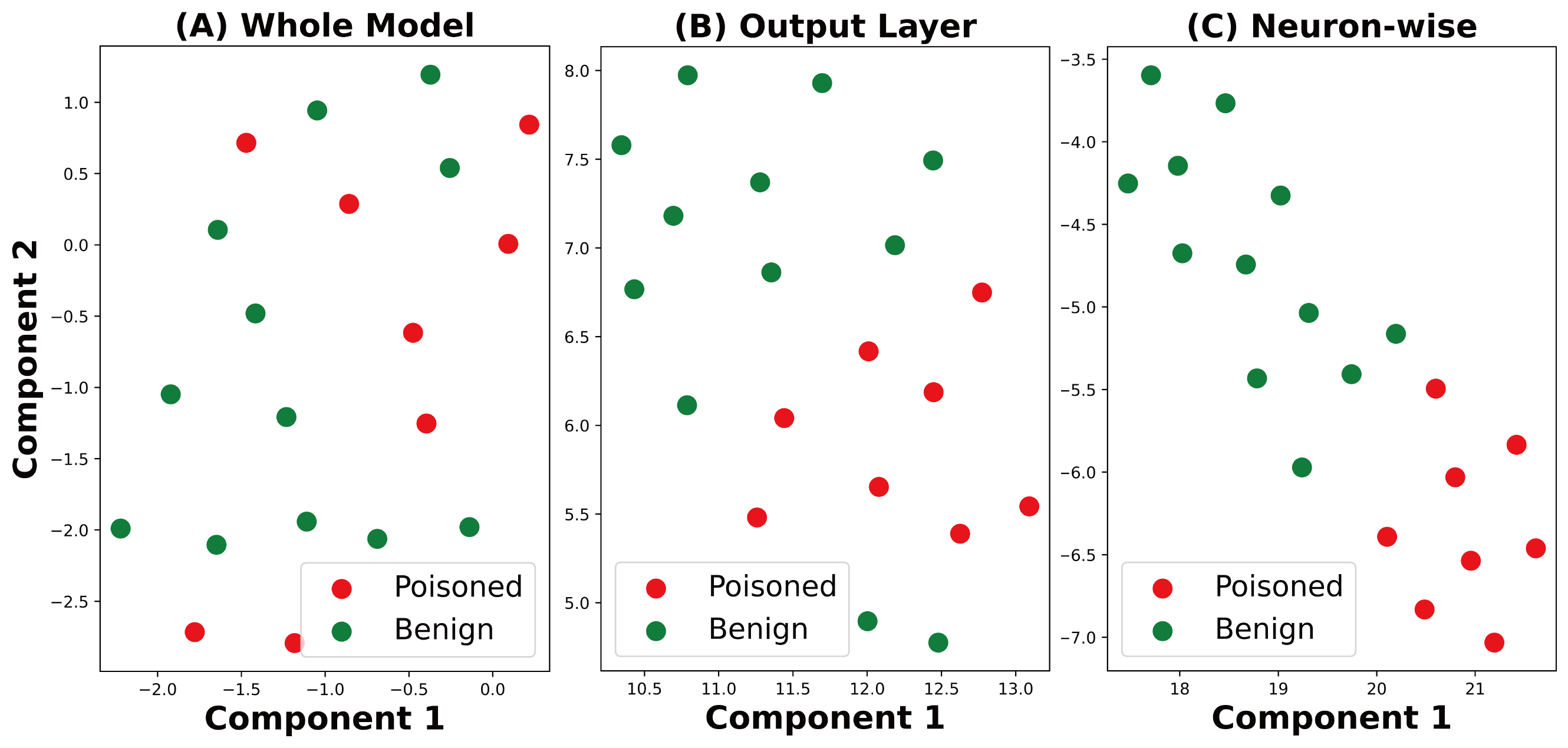}}
\caption{Comparison of poisoned and benign models based on three kinds of features: (A) whole model parameters, (B) output layer parameters, and (C) neuron-wise parameters (with two more distinctive clusters).}
\label{fig_scatter}
\end{figure}

\begin{figure}[tb]
\centerline{\includegraphics[width=0.48\textwidth]{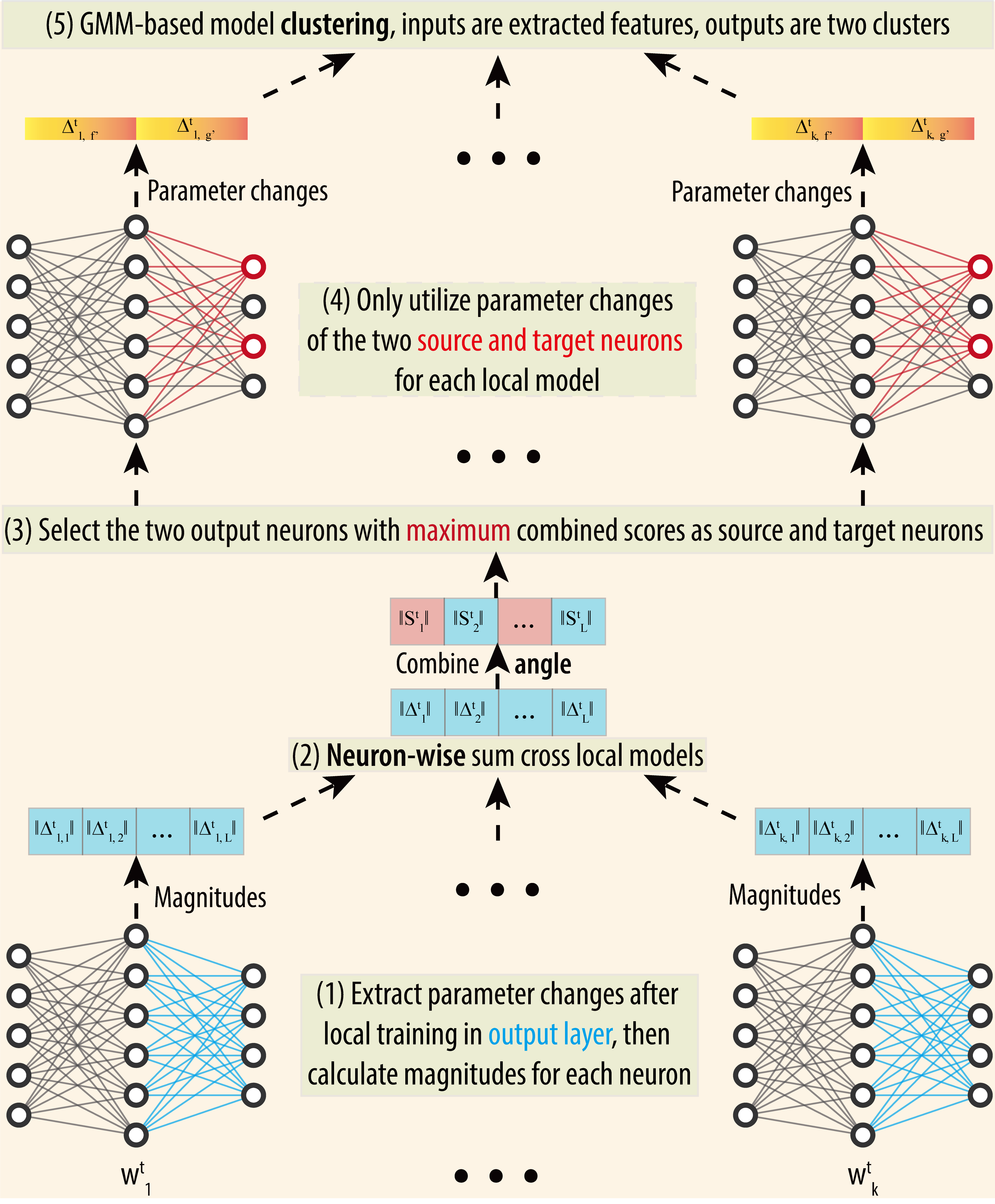}}
\caption{Workflow of poisoned local model detection in round $t$. Steps (1) and (4) extract neuron-wise features and are executed in parallel for each local model. Steps (2) and (3) identify source and target neurons in TLFAs. Step (5) executes the local model clustering based on GMM. Output layer parameters are marked in blue, while identified source and target neuron parameters are marked in red.}
\label{fig_DEFEND}
\end{figure}

\textbf{Intuition:} The detection module builds on the observation~\cite{jebreel2024lfighter} that the model parameters directly connected to the \textit{source and target neurons} exhibit the most pronounced discrepancies between benign and TLFA-poisoned updates. Intuitively, adversarial clients pursuing targeted label flipping optimize objectives that conflict with honest training, and this conflict is primarily captured by these neuron-adjacent parameters. Consequently, FedTrident detects TLFAs more efficiently by concentrating on this critical subset, rather than relying on the entire model (e.g., TMean~\cite{yin2018byzantine}) or the full output layer (e.g., FoolsGold~\cite{fung2020limitations}). Using the dimension reduction method Uniform Manifold Approximation and Projection (UMAP)~\cite{mcinnes2018umap}, Fig.~\ref{fig_scatter} illustrates that features derived from these neuron-wise parameters provide clearer separation between poisoned and benign models compared to alternative feature choices. Fig.~\ref{fig_DEFEND} depicts the detailed workflow of poisoned local model detection in round $t$.

\textbf{Feature Extraction:} Thus, in each FL round $t$ and for each output neuron $l \in L$, we define the client-specific parameter difference as Equation~(\ref{eq_delta}),

\begin{equation} \label{eq_delta}
\Delta^{t}_{k,l} = \omega^{t}_{k,l} - \omega^{t-1}_{l}
\end{equation}
where $\omega^{t}_{k,l}$ and $\omega^{t-1}_{l}$ are parameters associated with output neuron $l$ in the local and global models, respectively. We then compute the $\ell_2$-norm (magnitude) of $\Delta^{t}_{k,l}$ as Equation~(\ref{eq_client-neuron-norm}),

\begin{equation} 
\lVert \Delta^t_{k,l}\rVert_2 = \sqrt{\sum_{i=1}^{d_l} \big( \Delta^t_{k,l,i}\big)^2}, \qquad c_k\in\mathbb{C}^t; l\in L
\label{eq_client-neuron-norm}
\end{equation} 
where $d_l$ is the number of parameters associated with output neuron $l$. 

To identify the neurons most impacted by TLFAs, we first aggregate the magnitudes over all participating clients for each output neuron according to Equation~(\ref{eq_accumulated-neuron-magnitude}). 
\begin{equation} \lVert \Delta^t_{l}\rVert_2 = \sum_{c_k\in\mathbb{C}^t} \lVert \Delta^t_{k,l}\rVert_2, \qquad l\in L 
\label{eq_accumulated-neuron-magnitude} 
\end{equation}

Then, we compute the consensus direction (mean vector) for neuron $l$ as $\mu^t_l = \frac{1}{|\mathbb{C}^t|} \sum_{c_k \in \mathbb{C}^t} \Delta^t_{k,l}$. After that, we measure the angular inconsistency $I^t_l$ by calculating the average cosine similarity between each client's update and the consensus direction, as shown in Equation~(\ref{eq_angular_inconsistency}).

\begin{equation} \label{eq_angular_inconsistency}
I^t_l = 1 - \frac{1}{|\mathbb{C}^t|} \sum_{c_k \in \mathbb{C}^t} \frac{\Delta^t_{k,l} \cdot \mu^t_l}{\lVert \Delta^t_{k,l} \rVert_2 \lVert \mu^t_l \rVert_2}
\end{equation}
If the mean update vector is effectively zero ($\lVert \mu^t_l \rVert_2 \approx 0$), $I^t_l$ is set to $0$. A larger $I^t_l$ indicates a higher degree of directional disagreement among clients, which is a strong footprint of targeted poisoning attacks.

Finally, we compute a combined score $S^t_l$ that weights the accumulated magnitude by the angular inconsistency, as defined in Equation~(\ref{eq_combined_score}).

\begin{equation} \label{eq_combined_score}
S^t_l = \lVert \Delta^t_{l}\rVert_2 \times (1 + I^t_l)
\end{equation}

The two neurons with the highest combined scores $S^t_l$, denoted as $f^{t'}$ and $g^{t'}$, are selected, with $idx(f^{t'}) < idx(g^{t'})$, and they are interpreted as the source and target neurons in round $t$, respectively.

Next, we focus on the parameters directly connected to these two neurons, which are expected to encode the strongest attack footprint. Specifically, we construct the feature set $U^t$ according to Equation~(\ref{eq_fea_set}).

\begin{equation} 
\label{eq_fea_set}
U^t=\{\Delta^t_{k,l}|c_k \in \mathbb{C}^t, l \in \{f^{t'},g^{t'}\}\} 
\end{equation} 

\textbf{Clustering and Filtering:} We feed TLFA-specific features $U^t$ into a Gaussian Mixture Model (GMM) to partition client updates into two clusters, interpreted as one benign cluster $\mathbb{C}^t_{good}$ and one poisoned cluster $\mathbb{C}^t_{bad}$. Unlike hard clustering approaches, such as KMeans~\cite{jebreel2024lfighter} and HDBSCAN~\cite{LiuP:C:2025c}, GMM provides soft probabilistic assignments, avoiding brittle decision boundaries and better accommodating heterogeneity in client data and training dynamics (we consider this issue in both feature extraction and cluster algorithm design to ensure high detection accuracy). When the separation between benign and poisoned updates is ambiguous, GMM models the underlying parameter distributions to explicitly represent uncertainty, leading to more stable and calibrated detection. Following the empirical evidence in~\cite{jebreel2024lfighter} and the visualization in Fig.~\ref{fig_scatter}, the denser cluster is identified as the poisoned cluster, and its corresponding local models are discarded prior to global aggregation. 

\textbf{Model Validation:} As the source and target classes (i.e., the TLFA objective) in each round are inferred via neuron-wise analysis, we can further leverage such information to consistently track the robustness of the global model over training, using two critical metrics, SRE and ASR. If SRE decreases or ASR increases, beyond predefined thresholds $SRE^{\mathrm{thr}}$ or $ASR^{\mathrm{thr}}$,  compared to the previous round, the current global model is regarded as potentially compromised and is discarded. This validation step is optional and intended for deployments that can sustain additional computational overhead. Even without this independent validation, FedTrident remains effective in detecting poisoned models through its clustering-based filtering mechanism.  

\subsection{Malicious Vehicular Client Exclusion} 

To mitigate the long-term impact of persistent adversaries on the global model, FedTrident maintains a rating score $r_k \in [r^{min}, r^{max}]$ for each vehicular client $c_k$ based on the previous detection results. The score is updated at each round $t$ according to Equation~(\ref{eq_rating}),

\begin{equation}\label{eq_rating}
    r_k(t) = 
    \begin{cases}
    \displaystyle
    \min\bigl\{\,r_k(t-1) + \beta_k^t,\;r^{\mathrm{max}}\,\bigr\},  & \text{if $\omega_k^t$ good}\\[1ex]
    \displaystyle
    \max\bigl\{\,r_k(t-1) - \gamma_k^t,\;r^{\mathrm{min}}\,\bigr\}, & \text{if $\omega_k^t$ bad}
    \end{cases}
\end{equation}
where $\beta_k^t, \gamma_k^t \in (0, r^{max}]$ are values controlling the reward and penalty steps, respectively. By default, $\beta_k^t$ and $\gamma_k^t$ are constants, but they can also be adaptive, e.g., stage-wise.

Under this adaptive mechanism, $r_k$ decreases when $c_k$' updates are detected as poisoned in round $t$ and increases when they are deemed benign. By aggregating evidence over rounds, the scheme is robust to occasional detection noise or benign anomalies (e.g., limited local data and stochastic training variance), avoiding premature permanent exclusion. Once $r_k$ falls below $r^{min}$, client $c_k$ is labeled malicious with high confidence and added to blacklist, $\mathbb{B}$, thus removed from subsequent client selection, preventing sustained poisoning of the training process. This cumulative scoring reduces false positives while maintaining high sensitivity to persistent adversarial behavior.

As mentioned in the system model, we leverage standardized V2X security and privacy protocols in FedTrident, notably VPKI and pseudonyms, to ensure unlinkability, authenticity, and non-repudiation. With these V2X protocols, FedTrident can smoothly implement the malicious vehicular client exclusion strategy after the poisoned local model detection strategy.

\subsection{Corrupted Global Model Remediation}

The exclusion of a malicious client $c_a$ in round $T_{c_a}$ is not sufficient to remove its influence from the final global model $\omega^{T}$; the historical updates $\Delta \omega^t_{c_a}$ contributed by $c_a$ over rounds $t \in [1, T_{c_a}]$ should also be eliminated. In each round $t$, the global model $\omega^t$ could be seen by updating $\omega^{t-1}$ with the aggregated contribution from all participating clients. Let $\Delta \omega^t$ denote this aggregated update; then, the final global model can be expressed as the composition of the initial model $\omega^0$ and the sequence of updates from round 1 to $T$.

The direct strategy is to reconstruct an ``unlearned'' model by removing attacker $c_a$'s updates from the training trajectory. Due to the incremental nature of FL, each global model $\omega^t$ affects subsequent local training and aggregation. To analyze this dependency, we introduce $\varepsilon_t$ as a correction term at round $t$. If in each round, a constant number of clients, $M$, involved in aggregation, the initial approximation that excludes malicious client $c_a$'s contributions can be written as Equation~(\ref{eq_unlearning}).
\begin{equation}\label{eq_unlearning}
    \omega^{T_{c_a}^{\prime}}
    = \omega^0
    + \frac{M}{M-1}\sum_{t=1}^{T_{c_a}} \Delta \omega^t
    - \frac{1}{M-1}\sum_{t=1}^{T_{c_a}} \Delta \omega^t_{c_a}
    + \sum_{t=1}^{T_{c_a}} \varepsilon_t
\end{equation}

If $M$ is large, the factor $\tfrac{M}{M-1}$ is negligible for the aggregated updates $\Delta \omega^t$. The accumulated corrections $\varepsilon_t$ are also smaller if unlearning at the early round $T_{c_a}$ compared to unlearning at the final round $T$. Such that, the unlearning procedure could be simplified and approximated to Equation~(\ref{eq_unlearning_simp}), where $\Delta \omega^t_{c_a}=0$ if $c_a \notin \mathbb{C}^t$ (not selected as participant in round $t$) or $c_a \in \mathbb{C}^t_{bad}$ (selected as participant in round $t$ but having its contribution(s) filtered out of aggregation).

\begin{equation}
    \omega^{T_{c_a}^{\prime}}
    = \omega^{T_{c_a}}
    - \sum_{t=1}^{T_{c_a}}\frac{1}{|\mathbb{C}^t_{good}|} \Delta \omega^t_{c_a}
    \label{eq_unlearning_simp}
\end{equation}

Equation ~(\ref{eq_unlearning_simp}) requires to separately storing $\Delta \omega^t_{c_a}$ in every round $t$, as $|\mathbb{C}^t_{good}|$ varies over rounds. To further reduce complexity, we simplify Equation ~(\ref{eq_unlearning_simp}) as Equation ~(\ref{eq_unlearning_simp_2}),

\begin{equation}
    \omega^{T_{c_a}^{\prime}}
    = \omega^{T_{c_a}}
    - \frac{1}{M_{c_a}}\sum_{t=1}^{T_{c_a}} \Delta \omega^t_{c_a}
    \label{eq_unlearning_simp_2}
\end{equation}
where $M_{c_a}$ is the averaged $|\mathbb{C}^t_{good}|$ for those $t$ that $c_a \in \mathbb{C}^t_{good}$. In this way, only the accumulated historical update, $\Delta \omega^t_{k,sum}=\Delta \omega^{t-1}_{k,sum}+(\omega^t_{k}-\omega^{t-1})$, is stored for each client, instead of every individual raw update, $\Delta \omega^t_{k}=\omega^t_{k}-\omega^{t-1}$. Such that, Equation ~(\ref{eq_unlearning_simp_2}) could be rewritten as Equation ~(\ref{eq_unlearning_simp_3}).

\begin{equation}
    \omega^{T_{c_a}^{\prime}}
    = \omega^{T_{c_a}}
    - \frac{1}{M_{c_a}}\Delta \omega^{T_{c_a}}_{c_a,sum}
    \label{eq_unlearning_simp_3}
\end{equation}

In this way, we effectively and efficiently subtract the historical averaged contributions of attacker $c_a$ from the global model $\omega^t_{c_a}$ to suppress its influence.

\subsection{Complexity Analysis}

Let $d_w$, $d_o$, and $d_e$ denote the dimensionalities of the full DNN model, the output layer, and a single output neuron, respectively. In each round, the computational cost of FedTrident consists of:

\begin{enumerate}
    \item $\mathcal{O}(M d_o)$ to compute output-layer parameter deviations for $M$ clients;
    \item $\mathcal{O}(M E d_e)$ to calculate neuron-wise scores for $E$ output neurons across $M$ clients;
    \item $\mathcal{O}(E \log E)$ to recognize the source and target neurons from the $E$ candidates;
    \item $\mathcal{O}(M d_e)$ to perform GMM-based clustering on the neuron-wise features of $M$ clients;
    \item $\mathcal{O}(M)$ to update the blacklist $\mathbb{B}$, rating scores $r_k$, and accumulated historical updates $\Delta \omega^{t}_{k,\mathrm{sum}}$ involving $M$ clients.
\end{enumerate}

Given $d_w \gg d_o \gg d_e$ and $E$ bounded by the number of classes, the dominant term is $\mathcal{O}(M d_o)$, yielding the overall per-round complexity of FedTrident as $\mathcal{O}(M d_o)$, which is the same as state-of-the-art methods FLARE~\cite{LiuP:C:2025c} and DEFEND~\cite{LiuP:C:2025d}.

In contrast, defenses operating on the full model or pairwise client relations incur substantially higher costs, e.g., Median with $\mathcal{O}(M \log(M d_w))$, Trimmed Mean (TMean) with $\mathcal{O}(M \log(P d_w))$, Krum with $\mathcal{O}(M^2 d_w)$, and FoolsGold with $\mathcal{O}(M^2 d_o)$. Therefore, FedTrident is computationally efficient.

\subsection{Advantage Analysis} 

The neuron-wise analysis can identify the attack goal of TLFAs, even if \textit{dynamic adversaries} change their goals during the training. FedTrident takes full advantage of such information for critical feature extraction and model validation in each round, thus maximizing its mitigation effect. Moreover, our scheme readily \textit{generalizes to DNN-based autonomous driving tasks beyond RCC}, as the proposed neuron-wise analysis targets TLFAs at the output layer and is agnostic to the underlying classification semantics. Consequently, it can be applied to any DNN architecture that defines source–target label relationships. Finally, poisoned local model detection, malicious vehicular client exclusion, and corrupted global model remediation operate as an integrated, mutually reinforcing pipeline: 1) the \textit{detection} module provides round-wise, neuron-wise evidence of poisoned models; 2) the \textit{exclusion} module aggregates these outcomes over time to isolate persistently malicious clients; and 3) the \textit{remediation} module leverages both exclusion signals and historical contributions to repair compromised global models. Together, they ensure immediate attack suppression, long-term adversary containment, and final model integrity, forming a \textit{coherent FedTrident framework rather than three isolated modules}.

\section{Evaluation}\label{sec_eva}
\subsection{Experimental Settings}

We simulate an FL-RCC system with one server and $K=100$ vehicular clients implemented in PyTorch, running on an NVIDIA A100 Tensor Core GPU (40\,GB) and an Ice Lake CPU (16 cores, 128\,GB RAM). We also utilize an edge device suitable for vehicles, NVIDIA Jetson Orin Nano (1024-core Ampere GPU, 6-core Arm Cortex-A78AE CPU, and 8GB LPDDR5 RAM), to conduct real-world benchmarking experiments. In each round, the server randomly selects 20 clients (20\% participation), requesting them to perform local training. The default training configurations are listed in TABLE~\ref{tab_config}, and we build upon the FL framework of \cite{tolpegin2020data}\footnote{\url{https://github.com/git-disl/DataPoisoning_FL}}.

\begin{table}[t]
\centering
\caption{Default training configurations used in this paper.}
\label{tab_config}
\centering
\renewcommand{\arraystretch}{1.2} 
\begin{tabular}{p{0.21\textwidth}p{0.21\textwidth}}
\hline
\rowcolor{lightgray}
\textbf{Term}   &\textbf{Value}  \\  \hline
Loss Function      & Cross-Entropy \\
Batch Size         & 64                 \\
Learning Rate (LR) & 0.03              \\
Momentum for LR    & 0.5  \\
Optimizer          & SGD                \\
Local Epoch        & 3                \\
Total Round        & 60                \\
\hline
\end{tabular}
\end{table} 

\begin{figure}[tb]
\centerline{\includegraphics[width=0.48\textwidth]{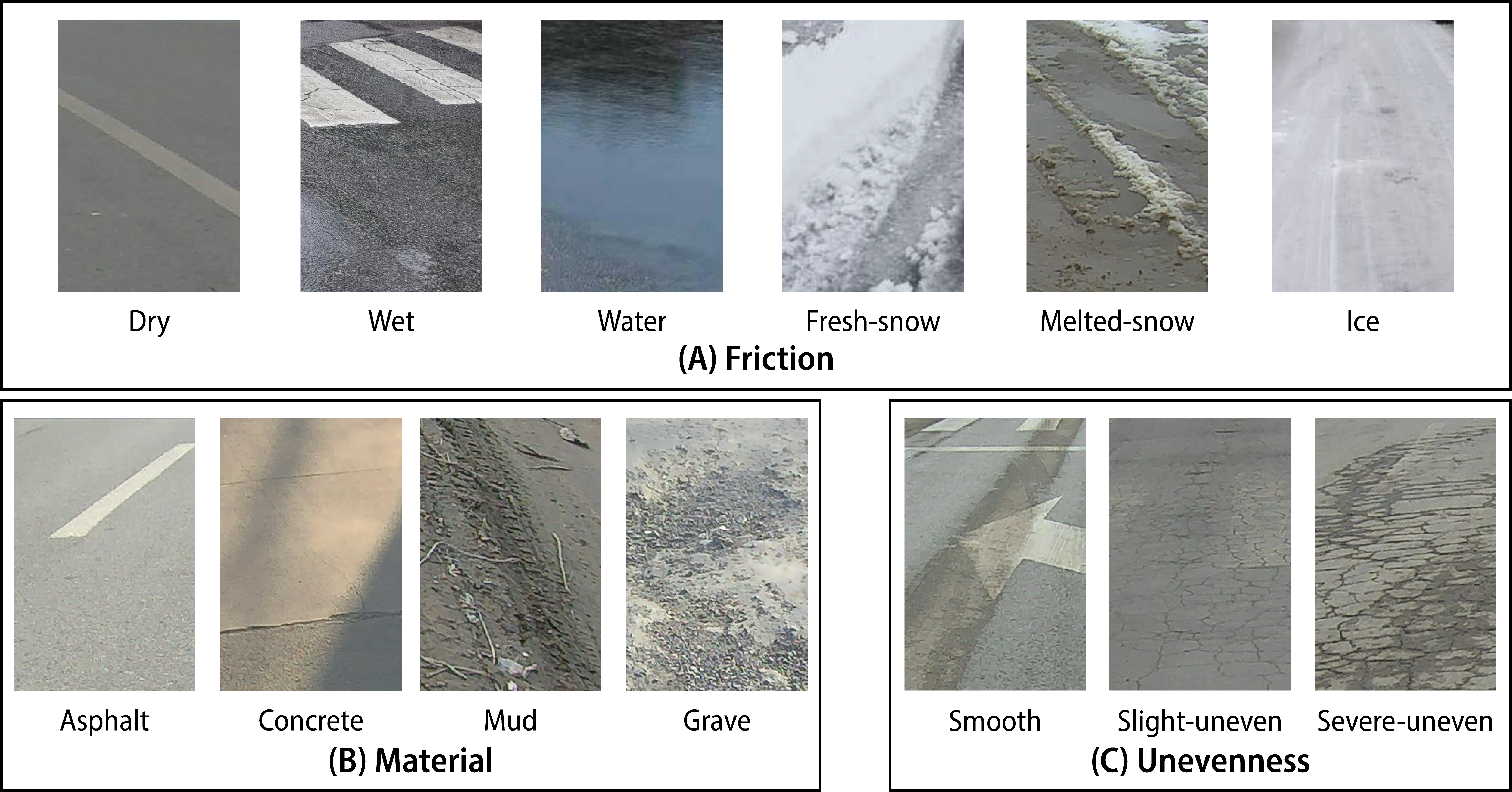}}
\caption{Image examples of RSCD dataset: (A) Friction level, (B) Material level, and (C) Unevenness level.}
\label{fig_rscd}
\end{figure}

\begin{figure}[tb]
\centerline{\includegraphics[width=0.48\textwidth]{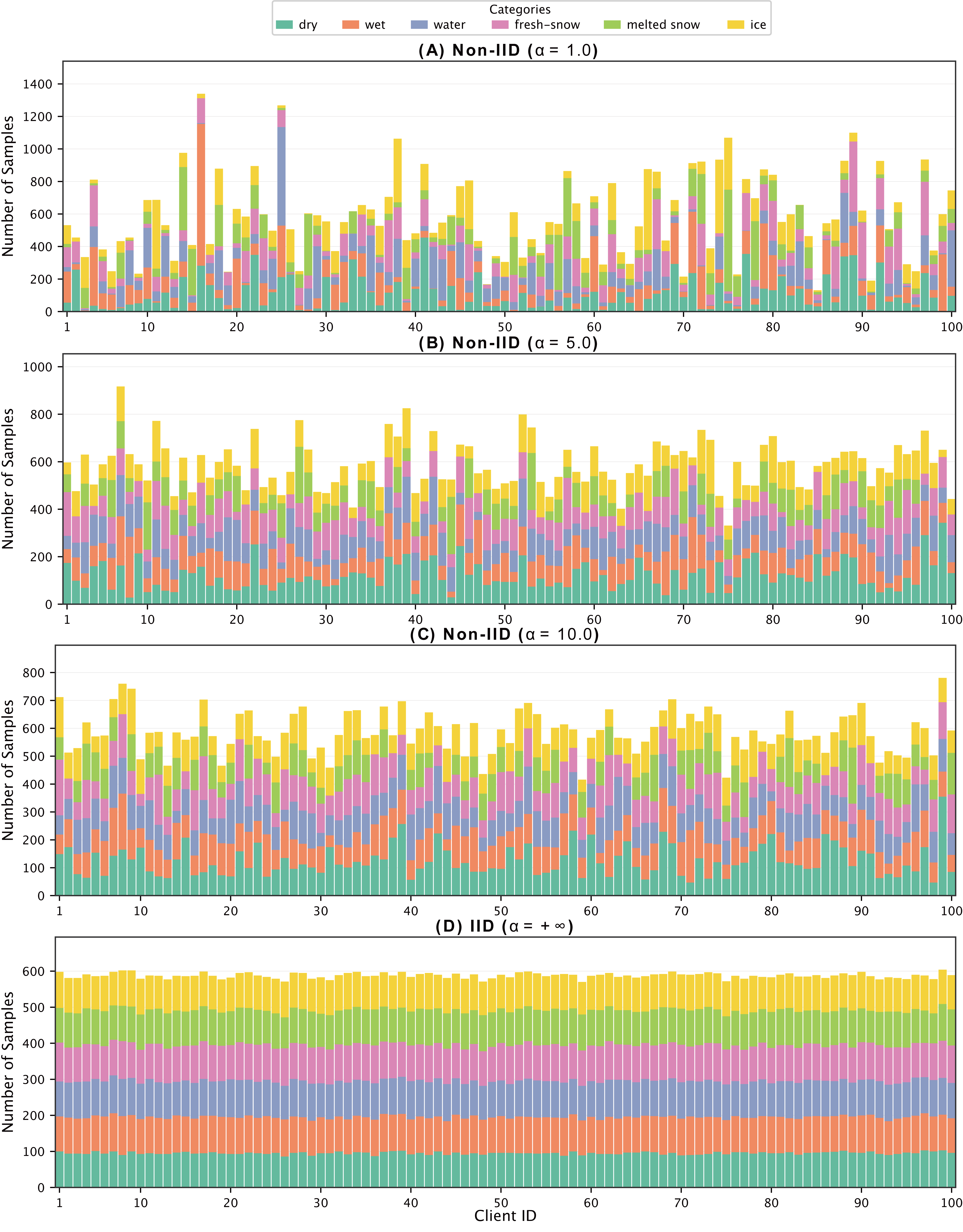}}
\caption{Local data distribution of 100 clients generated using the Dirichlet distribution with different $\alpha$ values based on RCC @ Friction dataset.}
\label{fig_distrib}
\end{figure}

\subsubsection{RCC Datasets and Tasks}\label{sec_task}

We utilize the Road Surface Classification Dataset (RSCD)\footnote{https://thu-rsxd.com/rscd/} for our evaluation. It contains 1 million real-world samples captured by vehicle-mounted cameras. We resize each image ($360\times240\times3$) to $224\times224\times3$ for efficiency. We create four data subsets; three of them are for three crucial RCC tasks, classification on a specific dimension, and the last one involves all three RCC tasks, representing a more complicated situation: 
\begin{itemize}
    \item \textit{RCC @ Friction}, which contains 58,800 training samples, 14,550 testing samples, and 6 labels (dry, wet, water, fresh-snow, melted-snow, and ice); 
    \item \textit{RCC @ Material}, which includes 57,000 training images and 15,000 testing images with 4 categories (asphalt, concrete, mud, and gravel); 
    \item \textit{RCC @ Unevenness}, which consists of 57,542 training pictures and 18,000 testing pictures labeled by smooth, slight-uneven, or severe-uneven; and 
    \item \textit{RCC @ All}, which comprises 54,000 training examples, 16,200 testing examples, and in total 27 classes. 
\end{itemize}

\begin{table}[t]
\centering
\caption{Summary of practical model information with both Jetson and A100, for the Friction task.}
\label{tab_model}
\centering
\renewcommand{\arraystretch}{1.2} 
\begin{tabular}{l l c c c}
\hline
\rowcolor{lightgray}
\textbf{Device}   &\textbf{Model}   &\textbf{Params}  &\textbf{Inference}  &\textbf{Memory}  \\
\rowcolor{lightgray}
\textbf{Type}    &\textbf{Type}    &\textbf{(FP32, M)}     &\textbf{Latency (ms)}  &\textbf{(Peak, M)}  \\\hline
\multirow{6}{*}{Jetson}
&ResNet-18                    & 42.65          &9.93   &60.97   \\
&ResNet-34                    & 81.21          &16.43  &101.07   \\
&MobileNet-V3                 &  5.81          &17.26  &16.53   \\
&EfficientNet-B1              & 24.88          &38.14  &43.87   \\
&DenseNet-121                 & 26.55          &50.79  &45.24   \\
&DeiT-Tiny                    & 21.08          &15.69  &32.10   \\
\hline
\multirow{6}{*}{A100}
&ResNet-18                    & 42.65          &2.18   &82.77   \\
&ResNet-34                    & 81.21          &4.03   &165.52   \\
&MobileNet-V3                 &  5.81          &4.36   &120.95   \\
&EfficientNet-B1              & 24.88          &9.62   &72.54   \\
&DenseNet-121                 & 26.55          &13.24  &93.82   \\
&DeiT-Tiny                    & 21.08          &4.21   & 32.10   \\
\hline
\end{tabular}
\end{table} 

Similar to the literature, e.g., \cite{jebreel2024lfighter}, this paper adopts the Dirichlet distribution to create Non-IID training data for each client. As illustrated in Fig.~\ref{fig_distrib}, a lower IID level $\alpha$ in the Dirichlet distribution represents a more heterogeneous data distribution ($\alpha=+\infty$ for IID); the default $\alpha$ is 1.0 in the experiments. By default, 30 of the 100 preselected clients are malicious (30\% malicious client rate). If included in the FL process, they execute the TLFAs by shifting the labels in their training datasets 1) from \emph{water} to \emph{dry} for Friction; 2) from \emph{gravel} to \emph{asphalt} for Material; 3) from \emph{severe-uneven} to \emph{smooth} for Unevenness, and 4) from \emph{wet-mud} to \emph{dry-asphalt-smooth} for RCC @ All. Testing datasets remain unchanged during training and are only used for inference.

\begin{table*}[tbp]
  \scriptsize
  \centering
  \caption{Overall results within default configurations on three RCC tasks. All values are ratios in \%.}
  \label{tab_results}
  \renewcommand{\arraystretch}{1.0} 
  \begin{tabular}{cc|cccc|cccc|cccc}
    \toprule
    \multicolumn{1}{c}{\multirow{2.5}{*}{Model Type}} &
    \multicolumn{1}{c|}{\multirow{2.5}{*}{Method}} &
    \multicolumn{4}{c|}{RCC @ Friction} &
    \multicolumn{4}{c|}{RCC @ Material} &
    \multicolumn{4}{c}{RCC @ Unevenness} \\
    \cmidrule(r){3-14}
    & &SRE$\uparrow$ &ASR$\downarrow$ &GAC$\uparrow$ &GAS$\uparrow$ &SRE$\uparrow$ &ASR$\downarrow$ &GAC$\uparrow$ &GAS$\uparrow$ &SRE$\uparrow$ &ASR$\downarrow$ &GAC$\uparrow$ &GAS $\uparrow$\\
    \midrule
    \multirow{10}{*}{ResNet-18} & FedAvg-NA$^{\ddagger}$\cite{mcmahan2017communication} & $72.28$ & $5.84$ & $85.26$ & $75.27$   & $82.24$& $3.07$ & $80.36$  & $71.84$    &$67.90$ &$9.93$ &$74.65$ &$66.65$ \\
    \cmidrule(r){2-14}
     & FedAvg\cite{mcmahan2017communication}  & $44.88$ & $30.85$ & $72.83$ & $63.23$   & $35.49$& $44.13$ & $70.17$ & $62.19$   & $46.74$ & $50.34$ & $69.14$ &$60.49$ \\
     & Krum\cite{blanchard2017machine}       & $48.89$ & $22.92$ & $81.44$ & $66.93$   & $23.09$& $21.84$ & $58.87$ & $48.66$   & $45.70$ & $53.38$ & $48.54$ & $39.93$ \\
     & TMean\cite{yin2018byzantine}      & $45.84$ & $18.48$ & $80.78$ & $68.54$  & $34.27$& $40.43$ & $55.75$ & $47.80$  & $44.37$ & $25.23$ & $52.65$ & $45.72$\\
     & Median\cite{yin2018byzantine}     & $48.44$ & $22.52$ & $81.56$ & $68.43$  & $46.85$& $25.07$ & $73.87$ & $64.89$  & $47.33$ & $34.55$ & $65.45$ & $58.19$  \\   
     \cite{he2016deep}& FoolsGold\cite{fung2020limitations}  & $50.64$ & $15.56$ & $81.64$ & $70.46$   & $42.53$& $35.20$ & $72.53$ & $63.26$   & $49.35$ & $26.33$ & $70.71$ & $60.83$ \\
     & FLAME\cite{nguyen2022flame}  & $51.20$ & $21.28$ & $63.32$ & $55.92$   & $40.66$ & $45.20$ & $52.51$ & $43.20$  & $16.29$ & $69.52$ & $41.36$ & $34.34$ \\
     & FLARE\cite{LiuP:C:2025c}  & $61.72$ & $14.64$ & $82.80$ & $72.19$   & $64.08$ & $14.69$ & $77.09$ & $68.92$  & $55.13$ & $28.13$ & $67.98$ & $60.02$ \\
     & DEFEND\cite{LiuP:C:2025d}  & $\underline{74.20}^\dagger$ & $\underline{2.84}$ & $\underline{84.93}$ & $\underline{72.76}$   & $\underline{80.11}$ & $\underline{2.88}$ & $\underline{78.61}$ & $\underline{69.43}$  & $\underline{79.85}$ & $\underline{7.32}$ & $\underline{73.71}$ & $\underline{62.33}$ \\
     \rowcolor{teal!12}
     \cellcolor{white}& FedTrident (Ours)      & $\textbf{88.04}^\ast$ & $\textbf{1.72}$ & $\textbf{85.16}$ & $\textbf{75.12}$  & $\textbf{87.47}$& $\textbf{1.89}$ & $\textbf{79.77}$ & $\textbf{71.76}$  & $\textbf{80.17}$ & $\textbf{5.68}$ & $\textbf{74.16}$ & $\textbf{67.21}$ \\
     \bottomrule
   \midrule

    \multirow{10}{*}{ResNet-34} & FedAvg-NA  & $72.12$ & $4.84$ & $85.74$ & $79.15$    & $83.92$& $3.07$ & $80.58$ & $73.46$    & $68.72$ & $5.87$ & $76.01$ & $69.18$ \\
    \cmidrule(r){2-14}
     & FedAvg        & $14.92$ & $56.96$ & $76.49$ & $70.41$  & $33.65$ & $41.44$ & $70.59$ & $62.78$  & $31.00$ & $40.62$ & $63.43$ & $57.39$ \\
     & Krum          & $23.76$ & $38.92$ & $66.56$ & $55.66$  & $45.28$ & $23.47$ & $65.00$ & $55.21$  & $16.13$ & $34.28$ & $55.14$ & $49.66$ \\
     & TMean         & $25.32$ & $44.76$ & $76.00$ & $68.03$  & $42.77$ & $32.37$ & $71.15$ & $62.85$  & $36.75$ & $28.05$ & $65.07$ & $58.63$ \\
     & Median        & $38.20$ & $26.68$ & $80.27$ & $73.57$  & $47.28$ & $25.04$ & $74.68$ & $66.60$  & $17.93$ & $33.03$ & $61.93$ & $57.21$ \\   
     \cite{he2016deep}& FoolsGold   & $43.84$ & $35.96$ & $80.84$ & $73.55$  & $44.48$ & $28.08$ & $74.39$ & $66.35$  & $34.18$ & $29.20$ & $66.94$ & $60.83$ \\
     & FLAME         & $41.28$ & $28.18$ & $62.23$ & $57.46$  & $27.57$ & $23.80$ & $59.03$ & $51.74$  & $\underline{61.85}$ & $\underline{12.28}$ & $62.11$ & $53.03$ \\
     & FLARE  & $57.40$ & $20.36$ & $82.99$ & $75.37$  & $57.63$ & $23.63$ & $75.39$ & $66.20$  & $50.40$ & $17.38$ & $68.97$ & $62.42$ \\
     & DEFEND  & $\underline{60.96}$ & $\underline{7.64}$ & $\underline{83.18}$ & $\underline{75.82}$   & $\underline{67.92}$ & $\underline{11.20}$ & $\underline{75.59}$ & $\underline{67.64}$  & $60.55$ & $14.73$ & $\underline{70.03}$ & $\underline{63.98}$ \\
     \rowcolor{teal!12}
     \cellcolor{white}& FedTrident (Ours)        & $\textbf{69.28}$ & $\textbf{2.36}$ & $\textbf{84.23}$ & $\textbf{77.61}$  & $\textbf{84.80}$& $\textbf{3.63}$ & $\textbf{79.77}$ & $\textbf{72.09}$  & $\textbf{73.25}$ & $\textbf{4.97}$ & $\textbf{73.86}$ & $\textbf{66.69}$ \\
    \bottomrule

   \midrule

    \multirow{10}{*}{MobileNet-V3} & FedAvg-NA  & $72.16$ & $5.40$ & $84.87$ & $77.23$   & $83.20$ & $2.93$ & $78.45$ & $69.70$   & $71.13$ & $6.90$ & $73.24$ & $66.50$ \\
    \cmidrule(r){2-14}
     & FedAvg        & $32.16$ & $40.96$ & $78.19$ & $70.45$   & $27.87$ & $46.71$ & $67.95$ & $60.31$   & $26.40$ & $43.75$ & $61.28$ & $55.32$ \\
     & Krum          & $25.84$ & $19.92$ & $59.50$ & $45.68$   & $36.05$ & $58.05$ & $45.26$ & $36.60$   & $8.05$ & $35.87$ & $49.38$ & $44.17$ \\
     & TMean         & $38.20$ & $33.24$ & $79.35$ & $71.14$   & $38.75$& $36.83$ & $69.39$ & $60.54$   & $45.08$ & $21.57$ & $67.95$ & $60.88$\\
     & Median        & $44.72$ & $27.48$ & $79.54$ & $70.64$   & $45.52$ & $24.59$ & $71.98$ & $63.54$   & $18.35$ & $36.87$ & $60.05$ & $54.83$  \\   
     \cite{howard2019searching}& FoolsGold   & $45.80$ & $28.84$ & $80.08$ & $71.85$   & $46.91$ & $28.53$ & $72.11$ & $63.22$   & $55.07$ & $20.28$ & $69.79$ & $62.31$ \\
     & FLAME  & $46.73$ & $19.84$ & $58.14$ & $45.98$   & $33.39$ & $22.59$ & $56.38$ & $48.21$   & $52.45$ & $19.95$ & $59.83$ & $50.88$ \\
     & FLARE  & $47.24$ & $21.92$ & $80.34$ & $72.07$   & $51.00$ & $20.08$ & $72.93$ & $64.35$   & $54.67$ & $16.18$ & $69.93$ & $62.50$ \\
     & DEFEND  & $\underline{63.40}$ & $\underline{7.24}$ & $\underline{81.80}$ & $\underline{74.07}$   & $\underline{67.92}$ & $\underline{9.76}$ & $\underline{74.47}$ & $\underline{66.23}$  & $\underline{61.87}$ & $\underline{13.67}$ & $\underline{71.16}$ & $\underline{63.64}$ \\
     \rowcolor{teal!12}
     \cellcolor{white}& FedTrident (Ours)  & $\textbf{76.96}$ & $\textbf{2.84}$ & $\textbf{82.56}$ & $\textbf{76.33}$   & $\textbf{85.28}$ & $\textbf{2.21}$ & $\textbf{77.93}$ & $\textbf{68.90}$   & $\textbf{77.93}$ & $\textbf{5.33}$ & $\textbf{71.59}$ & $\textbf{63.94}$ \\
    \bottomrule  

   \midrule

    \multirow{10}{*}{EfficientNet-B1} & FedAvg-NA  & $75.68$ & $3.40$ & $86.08$ &$74.18$    & $79.25$ & $4.75$ & $80.15$ &$73.50$     & $80.53$ & $8.08$ & $74.48$ &$67.30$ \\
    \cmidrule(r){2-14}
     & FedAvg      & $38.12$ & $32.96$ & $80.48$ &$68.44$    & $51.17$ & $25.20$ & $74.63$ &$65.83$    & $39.42$ & $29.42$ & $65.48$ &$57.66$   \\
     & Krum        & $4.92$  & $69.32$ & $64.49$ &$50.43$    & $42.88$ & $20.13$ & $52.64$ &$42.18$    & $43.67$ & $26.65$ & $66.98$ &$58.41$   \\
     & TMean       & $46.52$ & $27.04$ & $81.46$ &$69.09$    & $26.27$ & $47.39$ & $69.30$ &$58.44$    & $43.55$ & $26.78$ & $66.86$ &$58.37$  \\
     & Median      & $53.88$ & $22.16$ & $82.54$ &$69.91$    & $54.59$ & $20.19$ & $75.49$ &$66.01$    & $44.93$ & $28.02$ & $66.63$ &$58.72$ \\   
     \cite{tan2019efficientnet}& FoolsGold   & $59.80$ & $16.76$ & $83.64$ &$\underline{73.23}$   & $61.15$& $17.84$ & $77.84$ &$68.31$   & $38.32$ & $26.32$ & $66.48$ &$59.46$\\
     & FLAME  & $52.36$ & $20.96$ & $82.56$ &$71.32$   & $31.97$ & $14.32$ & $59.19$ &$50.48$   & $50.25$ & $20.15$ & $70.47$ &$62.33$ \\
     & FLARE  & $62.04$ & $16.08$ & $83.46$ &$72.17$    & $65.73$ & $12.88$ & $77.56$ &$69.69$   & $59.27$ & $17.65$ & $70.60$ &$62.78$ \\
     & DEFEND  & $\underline{80.40}$ & $\underline{5.40}$ & $\underline{84.43}$ & $73.14$   & $\underline{82.99}$ & $\underline{3.31}$ & $\underline{77.87}$ & $\underline{69.81}$  & $\underline{79.13}$ & $\underline{7.98}$ & $\underline{72.87}$ & $\underline{63.12}$ \\
     \rowcolor{teal!12}
     \cellcolor{white}& FedTrident (Ours)       & $\textbf{84.40}$ & $\textbf{3.92}$ & $\textbf{84.76}$ & $\textbf{75.38}$  & $\textbf{84.16}$& $\textbf{3.15}$ & $\textbf{79.50}$ & $\textbf{71.06}$  & $\textbf{80.75}$ & $\textbf{5.52}$ & $\textbf{72.91}$ & $\textbf{64.74}$ \\
    \bottomrule

   \midrule

    \multirow{10}{*}{DenseNet-121} & FedAvg-NA  & $77.96$ & $5.16$ & $86.18$ & $80.57$   & $77.09$ & $4.19$ & $81.73$ & $74.56$   & $73.13$ & $6.40$ & $76.71$ & $69.96$ \\
    \cmidrule(r){2-14}
     & FedAvg        & $29.72$ & $41.12$ & $78.93$ & $72.09$   & $33.20$ & $41.65$ & $71.63$ & $64.32$   & $15.70$ & $45.67$ & $60.38$ & $55.77$ \\
     & Krum          & $28.20$ & $28.04$ & $66.42$ & $54.09$   & $4.08$& $63.55$ & $54.65$ & $48.38$   & $6.17$ & $39.83$ & $51.50$ & $47.87$ \\
     & TMean         & $25.76$ & $37.84$ & $78.51$ & $72.14$   & $43.68$ & $35.60$ & $73.28$ & $64.99$   & $31.38$ & $26.00$ & $66.27$ & $60.50$ \\
     & Median        & $26.36$ & $31.52$ & $78.82$ & $72.53$   & $52.64$ & $33.52$ & $72.88$ & $63.57$   & $37.58$ & $23.92$ & $68.12$ & $62.03$ \\   
     \cite{huang2017densely}& FoolsGold   & $34.80$ & $36.52$ & $80.06$ & $73.37$   & $52.72$ & $24.35$ & $76.18$ & $67.93$   & $43.73$ & $24.70$ & $69.80$ & $63.29$ \\
     & FLAME  & $19.04$ & $23.36$ & $63.16$ & $52.77$   & $41.79$ & $16.33$ & $60.85$ & $52.78$   & $52.60$ & $17.23$ & $56.76$ & $47.82$ \\
     & FLARE  & $53.88$ & $20.72$ & $82.71$ & $75.53$   & $57.89$ & $16.32$ & $77.19$ & $68.78$   & $47.80$ & $17.05$ & $71.41$ & $64.90$ \\
     & DEFEND  & $\underline{59.52}$ & $\underline{10.64}$ & $\underline{83.36}$ & $\underline{76.34}$   & $\underline{71.76}$ & $\underline{11.60}$ & $\underline{78.57}$ & $\underline{71.44}$  & $\underline{56.50}$ & $\underline{11.43}$ & $\underline{72.42}$ & $\underline{66.94}$ \\
     \rowcolor{teal!12}
     \cellcolor{white}& FedTrident (Ours)  & $\textbf{74.68}$ & $\textbf{6.40}$ & $\textbf{85.12}$ & $\textbf{78.22}$   & $\textbf{85.04}$& $\textbf{2.37}$ & $\textbf{81.18}$ & $\textbf{73.61}$   & $\textbf{72.25}$ & $\textbf{4.10}$ & $\textbf{74.54}$ & $\textbf{67.74}$ \\
    \bottomrule

   \midrule

    \multirow{10}{*}{Deit-Tiny} & FedAvg-NA  & $77.32$ & $3.44$ & $86.54$ &$76.81$   & $80.61$ & $3.79$ & $80.31$ &$72.61$   & $78.80$ & $7.02$ & $74.08$ &$66.63$\\
    \cmidrule(r){2-14}
     & FedAvg      & $25.52$ & $48.40$ & $77.89$ &$63.94$   & $44.88$& $30.85$ & $72.83$ &$62.50$   & $30.03$ & $45.47$ & $60.93$ &$52.17$ \\
     & Krum        & $17.48$ & $38.04$ & $67.15$ &$59.31$   & $50.48$ & $31.81$ & $72.82$ &$62.72$    & $40.73$ & $27.60$ & $66.94$ &$57.34$ \\
     & TMean       & $34.88$ & $40.84$ & $64.03$ &$54.92$   & $41.76$ & $45.55$ & $47.37$ &$40.56$    & $41.73$ & $37.50$ & $64.93$ &$55.78$ \\
     & Median      & $33.64$ & $45.72$ & $78.91$ &$65.41$   & $47.33$ & $41.28$ & $51.81$ &$43.71$    & $55.72$ & $14.65$ & $\textbf{77.01}$ &$\textbf{66.25}$  \\   
     \cite{touvron2021training}& FoolsGold   & $53.60$ & $25.08$ & $82.61$ &$70.92$   & $65.23$ & $14.75$ & $\underline{76.85}$ &$\underline{65.92}$   & $45.23$ & $17.33$ & $68.59$ &$60.36$ \\
     & FLAME  & $45.60$ & $25.84$ &$57.54$ &$46.92$  & $43.57$ & $19.92$ & $42.36$ &$35.36$   & $49.47$ & $33.63$ & $60.34$ &$53.47$ \\
     & FLARE  & $58.84$ & $17.80$ & $83.02$ &$72.02$   & $52.27$ & $29.09$ & $72.19$ &$63.83$   & $53.70$ & $18.88$ & $68.77$ &$61.20$ \\
     & DEFEND  & $\underline{80.76}$ & $\underline{4.84}$ & $\underline{84.58}$ & $\underline{72.76}$   & $\underline{82.11}$ & $\underline{4.75}$ & $65.55$ & $65.03$  & $\underline{78.52}$ & $\underline{7.63}$ & $69.14$ & $62.37$ \\
     \rowcolor{teal!12}
     \cellcolor{white}& FedTrident (Ours)    & $\textbf{87.84}$ & $\textbf{1.00}$ & $\textbf{85.54}$ & $\textbf{78.12}$  & $\textbf{85.81}$& $\textbf{2.48}$ & $\textbf{78.23}$ & $\textbf{69.60}$  & $\textbf{82.42}$ & $\textbf{2.33}$ & $\underline{72.69}$ & $\underline{65.33}$ \\
    \bottomrule
  
  \end{tabular}
\begin{tablenotes} 
	\footnotesize
	\item$^\ast$ \textbf{Bold numbers are the best performance.}
	\item$^\dagger$ \underline{Numbers with underline are the second-best values.} 
    \item$^{\ddagger}$ NA denotes No Attack. Others without this symbol are all under attack.
\end{tablenotes}
\end{table*}

\subsubsection{RCC Models}\label{sec_aimodel}

To evaluate the generality and compatibility of each method, we train six popular DNN models under FL for the RCC tasks:

\begin{itemize}
    \item \textit{ResNet-18} and \textit{ResNet-34}~\cite{he2016deep}: the residual framework can better support deeper networks.
    \item \textit{MobileNet-V3}~\cite{howard2019searching}: the architecture is suitable for mobile and resource-constrained environments.
    \item \textit{EfficientNet-B1}~\cite{tan2019efficientnet}: the model can scale depth, width, and resolution uniformly.
    \item \textit{DenseNet-121}~\cite{huang2017densely}: the dense convolutional network can adopt shorter connections between layers.
    \item \textit{Deit-Tiny}~\cite{touvron2021training}: the data-efficient image transformer model can achieve high performance through attention.
\end{itemize}

We choose these models because they are all lightweight DNN versions, thus suitable for vehicles. The detailed model size (MB), inference latency (ms), and peak memory (MB) of each model on both Jetson and A100 are summarized in TABLE~\ref{tab_model}. It clearly shows that the current edge device can support RCC tasks in real time.

\begin{figure*}[tbp]
\centerline{\includegraphics[width=0.97\textwidth]{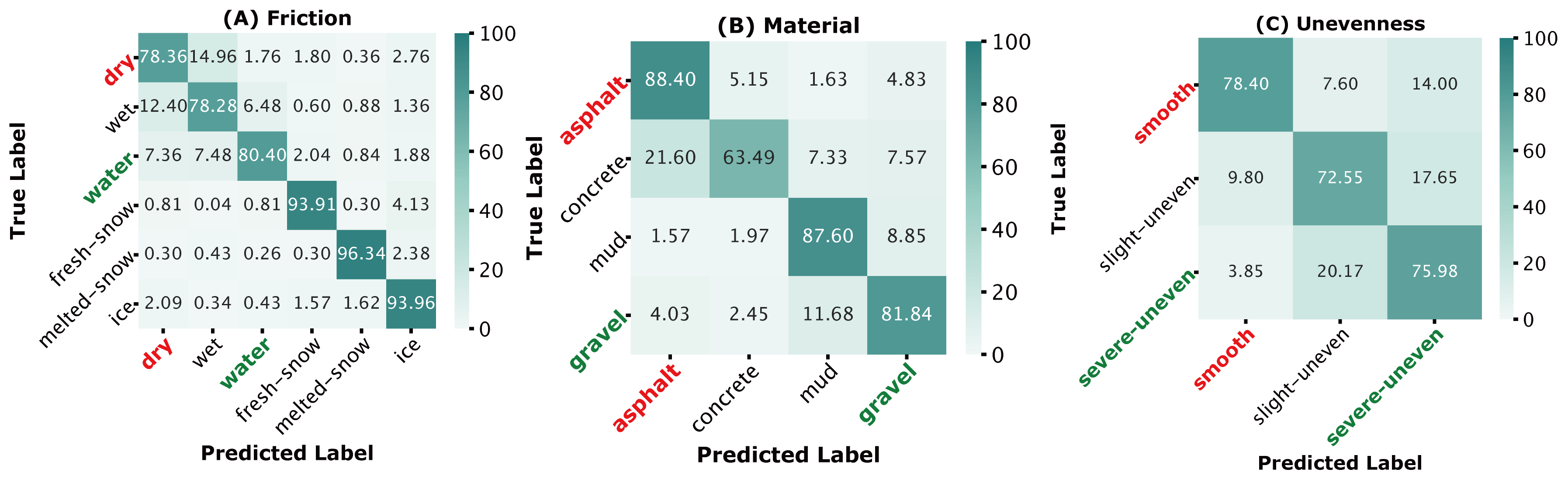}}
\vspace{-1.0em}
\caption{Confusion matrices of FedTrident with ResNet-18 in three RCC tasks. Classes in green and red are \textbf{\textcolor{mplGreen}{source}} and \textbf{\textcolor{mplRed}{target}} classes, respectively.}
\label{fig_confusion_resnet}

\centerline{\includegraphics[width=0.97\textwidth]{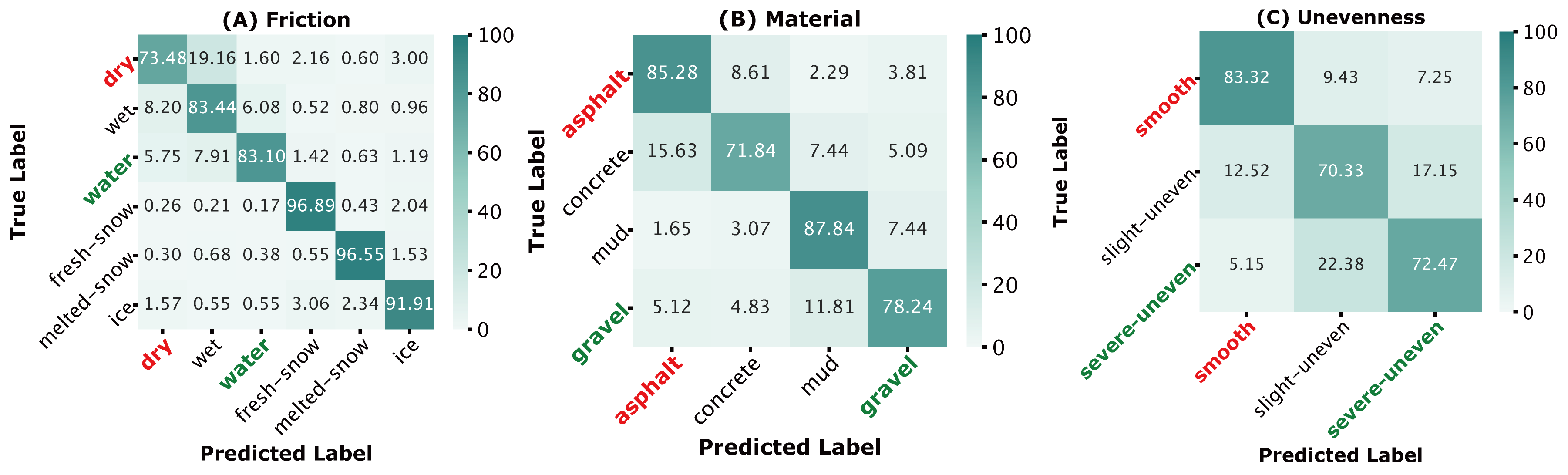}}
\vspace{-1.0em}
\caption{Confusion matrices of FedTrident with EfficientNet-B1 in three RCC tasks. Classes in green and red are \textbf{\textcolor{mplGreen}{source}} and \textbf{\textcolor{mplRed}{target}} classes, respectively.}
\label{fig_confusion_efficientnet}

\centerline{\includegraphics[width=0.97\textwidth]{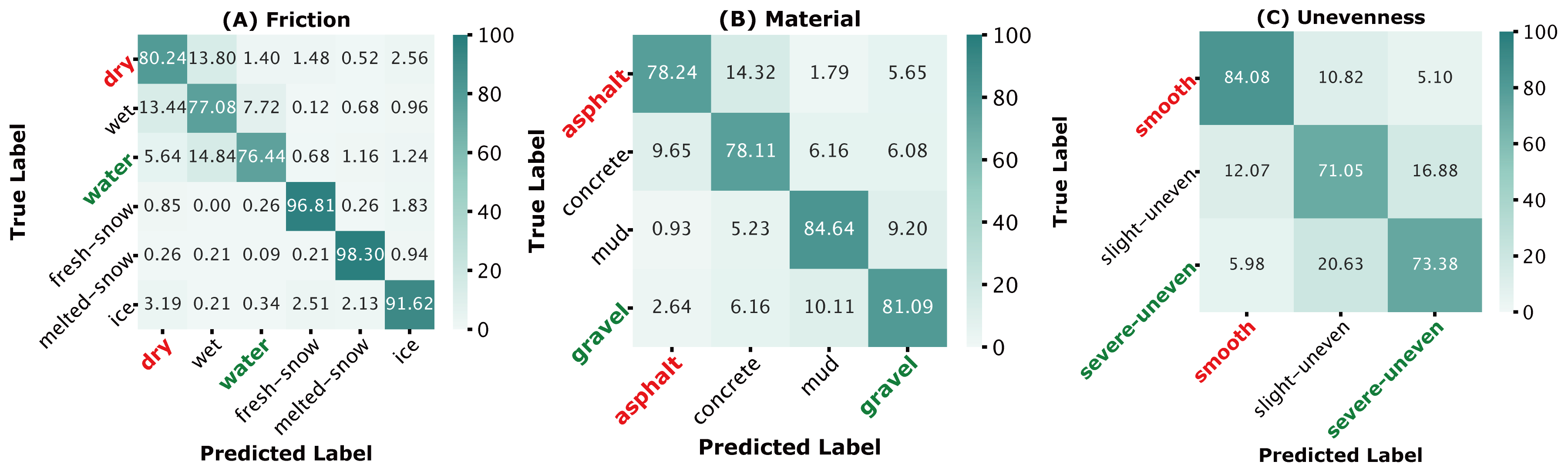}}
\vspace{-1.0em}
\caption{Confusion matrices of FedTrident with DeiT-Tiny in three RCC tasks. Classes in green and red are \textbf{\textcolor{mplGreen}{source}} and \textbf{\textcolor{mplRed}{target}} classes, respectively.}
\label{fig_confusion_deit}
\end{figure*}

\begin{figure}[tbp]
\centering 
\includegraphics[width=0.485\textwidth]{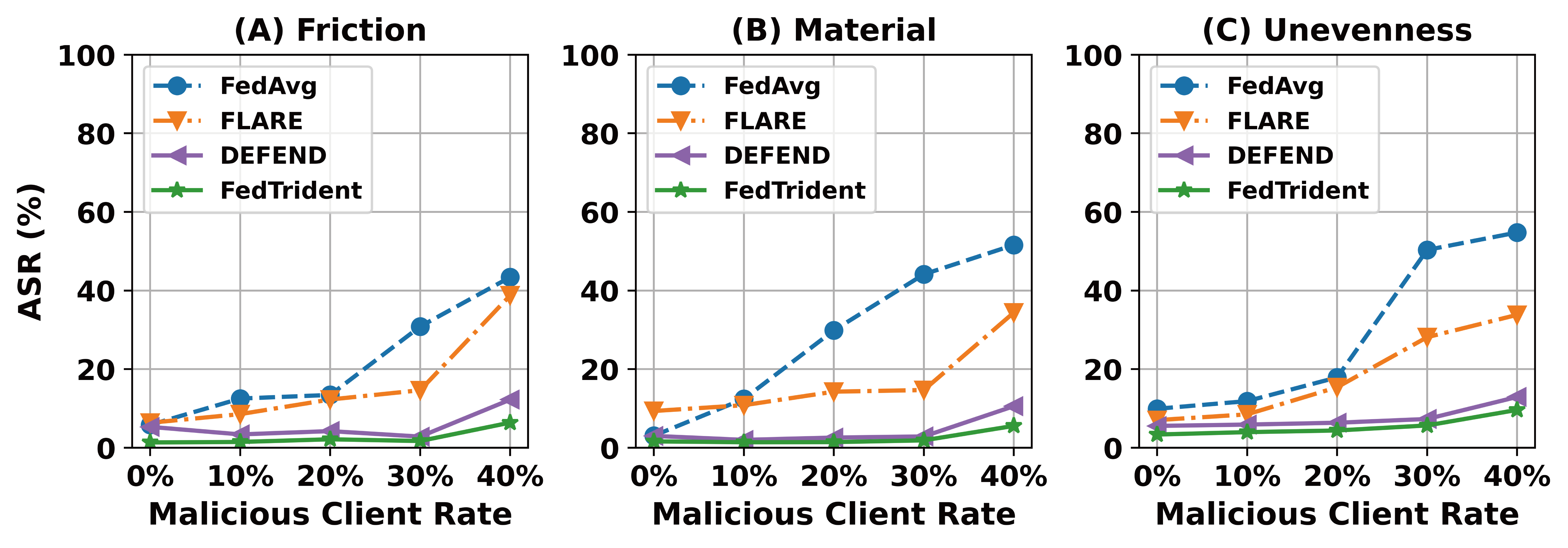}
\caption{ASR curves of malicious client rates with ResNet-18.}
\label{fig_poi_resnet}

\includegraphics[width=0.485\textwidth]{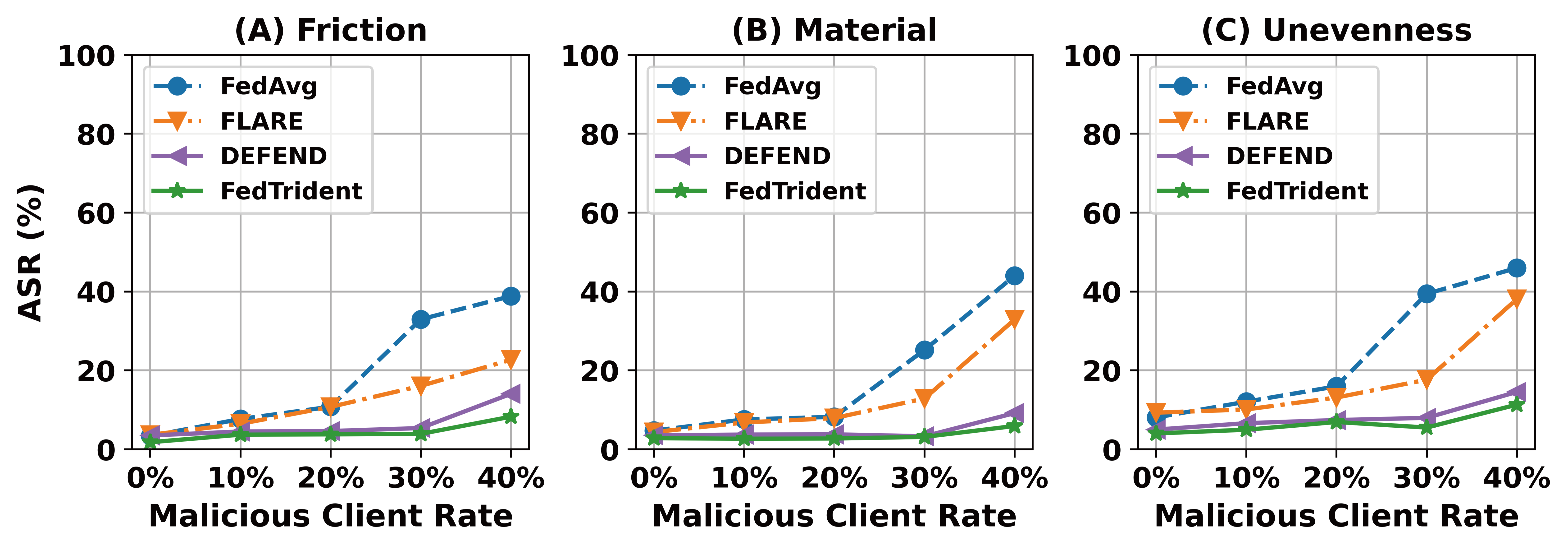}
\caption{ASR curves of malicious client rates with EfficientNet-B1.}
\label{fig_poi_efficientnet}

\includegraphics[width=0.485\textwidth]{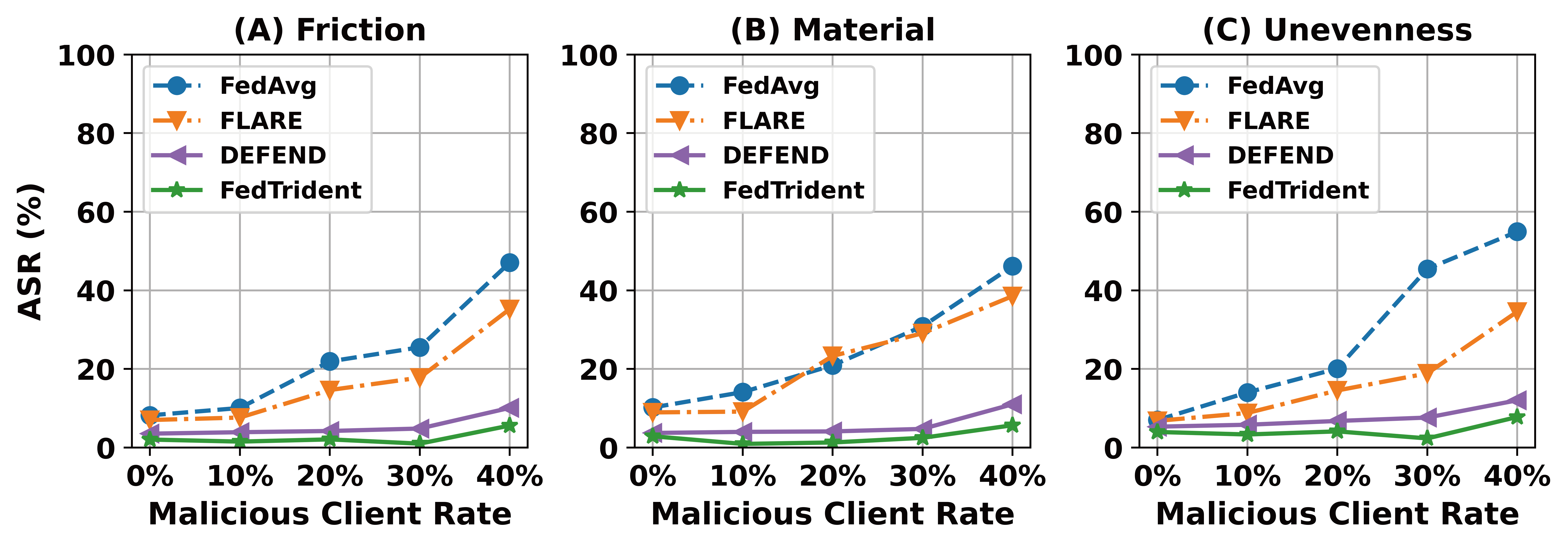}
\caption{ASR curves of malicious client rates with DeiT-Tiny.}
\label{fig_poi_deit}
\end{figure}

\subsubsection{Compared Methods}\label{sec_compared_method}

We compare the following eight methods in the experiments: FedAvg~\cite{mcmahan2017communication},  Krum~\cite{blanchard2017machine}, TMean~\cite{yin2018byzantine}, Median~\cite{yin2018byzantine}, FoolsGold~\cite{fung2020limitations}, FLAME~\cite{nguyen2022flame}, FLARE~\cite{LiuP:C:2025c}, and our proposed FedTrident ($r^{max}=1.00$, $r^{min}=0.00$, $r_k(0)=0.80$, $\beta_k^t=0.05$, $\gamma_k^t=0.15\times NC_k^t$, and $SRE^{thr}/ASR^{thr}=0.1$, where $NC_k^t$ is the current number of times it has been continuously identified as poisoned).

\subsubsection{Evaluation Objectives and Metrics}

We evaluate FedTrident, in comparison to the baseline methods (Section~\ref{sec_compared_method}) in terms of (i) detection effectiveness; (ii) attack mitigation; (iii) robustness against a gamut of adversaries, notably including dynamic ones (Section~\ref{sec_adv_model}); and (iv) generalization based on multiple models (Section~\ref{sec_aimodel}) and tasks (Section~\ref{sec_task}. The following four evaluation metrics quantify different performance aspects:

\begin{itemize}

    \item \textit{SRE}: Source Recall, the proportion of the correct predictions for the source class $s_{idx(f)}$ to the total number of samples in the source class computed by Equation~(\ref{eq_srec}).
    
\begin{equation}\label{eq_srec}
    SRE = \frac{n_{idx(f),idx(f)}}{\sum\limits_{i=1}^{E}n_{idx(f),i}}
\end{equation}  
    
    \item \textit{ASR}: Attack Success Rate, the ratio of samples with the source label $s_{idx(f)}$ misclassified into the target class $s_{idx(g)}$ based on Equation~(\ref{eq_asr}).

\begin{equation}\label{eq_asr}
    ASR = \frac{n_{idx(f),idx(g)}}{\sum\limits_{i=1}^{E}n_{idx(f),i}}
\end{equation} 

    \item \textit{GAC}: Global Accuracy, the ratio of all correct predictions to the total number of testing instances calculated by Equation~(\ref{eq_acc}).
    
\begin{equation}\label{eq_acc}
    GAC = \frac{\sum\limits_{i=1}^{E}n_{i,i}}{\sum\limits_{i=1}^{E}\sum\limits_{j=1}^{E}\times n_{i,j}}
\end{equation}     

    \item \textit{GAS}: Global Accuracy of Safety, defined by Equation~(\ref{eq_error_new}).
\end{itemize}


\subsection{Overall Results Analysis}\label{sec_result_anal}

\subsubsection{\textbf{Key Findings}} 
FL-RCC is vulnerable to TLFAs (Section~\ref{sec_tlfa_inf}), and the state-of-the-art is insufficient in thwarting TLFAs (Section~\ref{sec_baseline_perf}). FedTrident not only outperforms defense baselines but also shows no performance gap compared to FedAvg-NA, i.e., it achieves under-attack detection and mitigation that lead to model performance equivalent to that of FL without attack (Section~\ref{sec_fedtrident_perf}). It is also stable and compatible across tasks and models (Section~\ref{sec_fedtrident_perf}). Moreover, FedTrident remains robust even against a high rate of adversarial clients, even when they dynamically change their attack pattern (Section~\ref{sec_resil_fedtrident}). We elaborate on these findings in the rest of this section.

\subsubsection{\textbf{TLFA Influence}}\label{sec_tlfa_inf}
The TLFA influence on FL-RCC can be analyzed based on the results of FedAvg and FedAvg-NA; NA denotes ``No Attack" and acronyms for methods without NA imply the method is under TLFAs in TABLE~\ref{tab_results}. Specifically, if no defense is implemented, TLFAs reduce SRE by 42.95\%, increase ASR by 35.69\%, decrease GAC by 9.57\%, and drop GAS by 9.99\% on average. \textit{Such significant performance deterioration (notably in SRE and ASR) shows the vulnerability of FL-RCC to TLFAs}.

\subsubsection{\textbf{Baseline Performance}}\label{sec_baseline_perf} 
As per TABLE~\ref{tab_results}, the effectiveness of most of the baseline methods (including FedAvg, Krum, TMean, Median, FoolsGold, FLAME, and FLARE) is limited; even when comparing the best among them in a group (that is, evaluations for a specified model-task combination) to FedAvg-NA, the average gap is 18.43\%, 12.26\%, 3.56\%, and 4.49\% for SRE, ASR, GAC, and GAS, respectively. \textit{Such observations of innegligible performance gap manifest that most existing solutions are insufficient in thwarting TLFAs in FL-RCC}. As for the best and most recent baseline DEFEND, it is still worse than FedAvg-NA by 4.75\%, 2.82\%, 3.51\% and 3.79\% for the four metrics, respectively.

\begin{figure}[tbp]
\centerline{\includegraphics[width=0.485\textwidth]{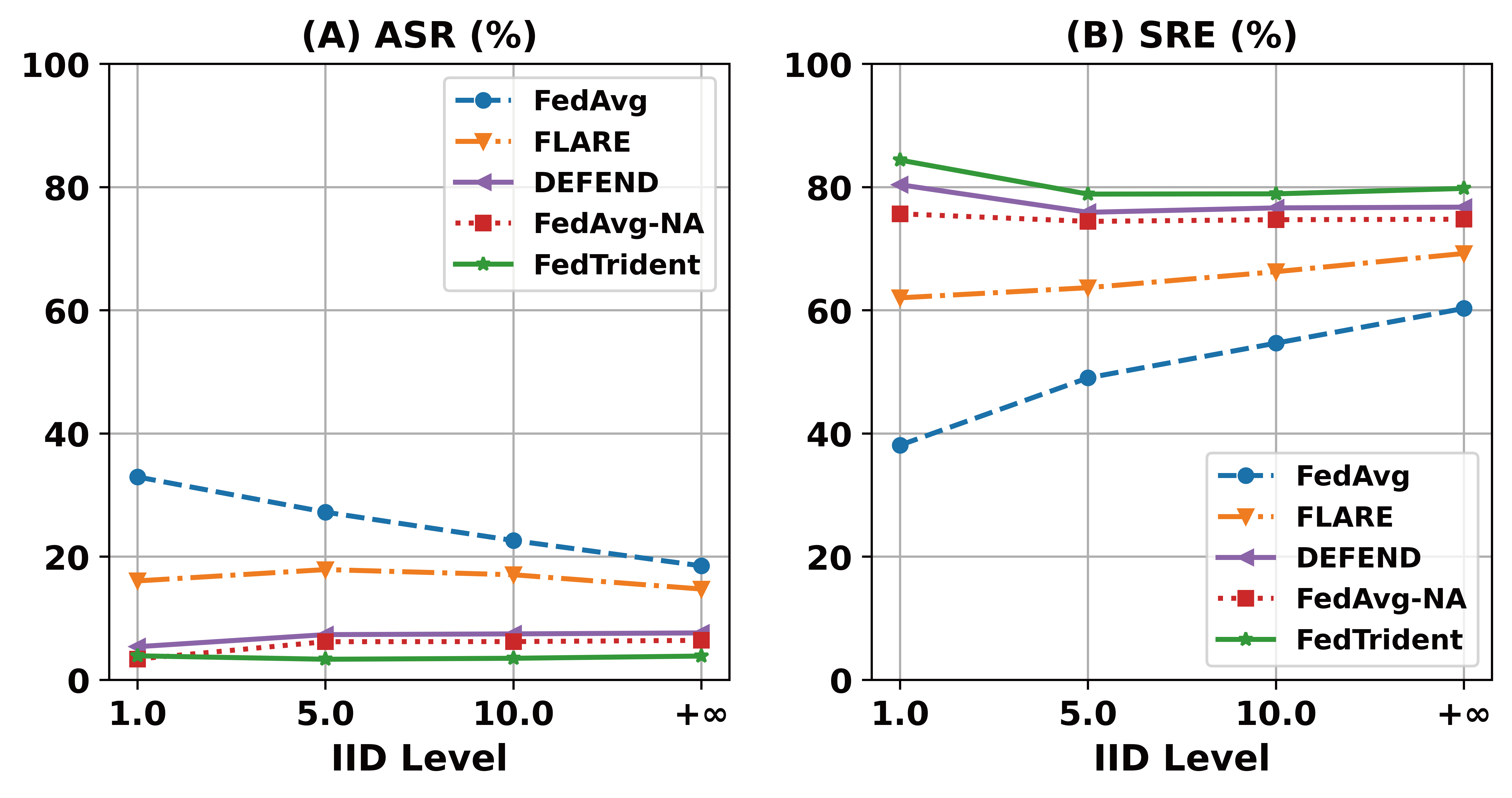}}
\caption{ASR and SRE curves of IID levels with EfficientNet-B1 in Friction.}
\label{fig_noniid_efficientnet}
\end{figure}

\subsubsection{\textbf{FedTrident Performance}}\label{sec_fedtrident_perf} 
Our solution, FedTrident, is the \textit{best} countermeasure in all 18 groups in terms of the two most important metrics, SRE and ASR. Notably, compared to the best baseline, the average improvements are 9.49\%, 4.47\%, 1.23\%, and 2.32\% for SRE, ASR, GAC, and GAS, respectively. Moreover, the average SRE value over groups is 81.14\%, ASR value is 3.44\%, GAC value is 79.08\%, and GAS value is 71.30\%, which are very close to the performance of FedAvg-NA (76.34\%, 5.23\%, 80.30\%, and 72.51\%). FedTrident is even better than FedAvg-NA in terms of SRE and ASR. \textit{This performance boost over the defense baselines with no gap to the FedAvg-NA demonstrates the superiority and effectiveness of FedTrident}.

\begin{table}[tbp]
  \scriptsize
  \centering
  \caption{Results on RCC @ All dataset. All values are ratios in \%.}
  \label{tab_results_27}
  \renewcommand{\arraystretch}{1.0} 
  \begin{tabular}{cc|cccc}
    \toprule
    Model Type &Method  & SRE$\uparrow$& ASR$\downarrow$ & GAC$\uparrow$ & GAS$\uparrow$ \\
    \midrule
    \multirow{10}{*}{ResNet-18} & FedAvg-NA$^{\ddagger}$\cite{mcmahan2017communication}  & $73.17$ & $0.17$ & $60.64$ & $27.03$ \\
    \cmidrule(r){2-6}
     & FedAvg\cite{mcmahan2017communication}    & $42.00$ & $30.17$ & $59.49$ & $25.82$  \\
     & Krum\cite{blanchard2017machine}          & $45.33$ & $25.33$ & $50.27$ & $21.61$   \\
     & TMean\cite{yin2018byzantine}             & $46.17$ & $22.50$ & $59.78$ & $26.14$ \\
     & Median\cite{yin2018byzantine}            & $59.83$ & $24.00$ & $59.60$ & $26.22$   \\   
     \cite{he2016deep}& FoolsGold\cite{fung2020limitations} & $45.67$ & $24.83$ & $59.75$ & $26.17$ \\
     & FLAME\cite{nguyen2022flame}              & $49.94$ & $23.71$ & $59.59$ & $26.03$   \\
     & FLARE\cite{LiuP:C:2025c}                 & $58.83$ & $19.33$ & $59.85$ & $26.26$   \\
     & DEFEND\cite{LiuP:C:2025d}                 & $\underline{68.50}^\dagger$ & $\underline{9.00}$ & $\underline{59.90}$ & $\underline{26.56}$   \\
     \rowcolor{teal!12}
     \cellcolor{white}& FedTrident (Ours)       & $\textbf{74.83}^\ast$ & $\textbf{0.00}$ & $\textbf{60.43}$ & $\textbf{26.69}$ \\
     \bottomrule
   \midrule

    \multirow{10}{*}{EfficientNet-B1} & FedAvg-NA & $77.17$ & $0.33$ & $64.10$ & $28.59$   \\
    \cmidrule(r){2-6}
     & FedAvg     & $26.17$ & $42.83$ & $62.07$ & $25.97$   \\
     & Krum       & $44.83$ & $27.50$ & $50.22$ & $22.30$   \\
     & TMean      & $51.83$ & $26.17$ & $\textbf{63.09}$ & $27.31$   \\
     & Median     & $51.00$ & $22.67$ & $62.75$ & $27.30$   \\   
     \cite{tan2019efficientnet}& FoolsGold  & $54.00$ & $23.33$ & $\underline{62.77}$ & $27.46$ \\
     & FLAME      & $53.33$ & $22.55$ & $62.44$ & $27.40$   \\
     & FLARE      & $59.17$ & $19.50$ & $62.04$ & $27.32$  \\
     & DEFEND     & $\underline{68.67}$ & $\underline{11.67}$ & $62.19$ & $\underline{27.47}$   \\
     \rowcolor{teal!12}
     \cellcolor{white}& FedTrident (Ours)      & $\textbf{77.17}$ & $\textbf{0.67}$ & $61.75$ & $\textbf{27.51}$  \\
    \bottomrule
   \midrule

    \multirow{10}{*}{Deit-Tiny} & FedAvg-NA & $75.50$ & $0.33$ & $64.06$ & $28.43$ \\
    \cmidrule(r){2-6}
     & FedAvg     & $23.67$ & $39.50$ & $61.98$ & $27.09$  \\
     & Krum       & $42.50$ & $31.45$ & $48.59$ & $20.91$  \\
     & TMean      & $55.83$ & $18.67$ & $62.84$ & $27.70$  \\
     & Median     & $46.17$ & $25.33$ & $61.59$ & $26.77$  \\   
     \cite{touvron2021training}& FoolsGold  & $47.67$ & $20.83$ & $\underline{62.76}$ & $27.65$ \\
     & FLAME      & $42.11$ & $32.76$ & $58.24$ & $27.73$ \\
     & FLARE      & $49.17$ & $16.83$ & $60.94$ & $26.70$  \\
     & DEFEND     & $\underline{61.50}$ & $\underline{9.83}$ & $62.43$ & $\underline{27.81}$   \\
     \rowcolor{teal!12}
     \cellcolor{white}& FedTrident (Ours)   & $\textbf{71.17}$ & $\textbf{1.17}$ & $\textbf{63.14}$ & $\textbf{27.96}$ \\
    \bottomrule
  
  \end{tabular}
\begin{tablenotes} 
	\footnotesize
	\item$^\ast$ \textbf{Bold numbers are the best performance.}
	\item$^\dagger$ \underline{Numbers with underline are the second-best values.} 
    \item$^{\ddagger}$ NA denotes No Attack. Others without this symbol are all under attack.
\end{tablenotes}
\end{table}

\begin{table}[tbp]
  \scriptsize
  \centering
  \caption{Results with dynamic attacks based on EfficientNet-B1. All values are ratios in \%.}
  \label{tab_results_dynamic}
  \renewcommand{\arraystretch}{1.0} 
  \begin{tabular}{cc|cccc}
    \toprule
    RCC Task &Method & SRE$\uparrow$& ASR$\downarrow$ & GAC$\uparrow$ & GAS$\uparrow$ \\
    \midrule
    \multirow{8}{*}{Friction} 
     & FedAvg\cite{mcmahan2017communication}    & $37.20$ & $36.88$ & $78.98$ & $70.55$  \\
     & Krum\cite{blanchard2017machine}          & $3.12$  & $67.36$ & $64.52$ & $55.99$   \\
     & TMean\cite{yin2018byzantine}             & $52.20$ & $27.04$ & $82.25$ & $74.47$ \\
     & Median\cite{yin2018byzantine}            & $49.08$ & $23.36$ & $81.40$ & $73.34$   \\   
     & FoolsGold\cite{fung2020limitations}      & $51.88$ & $20.56$ & $82.52$ & $74.91$ \\
     & FLAME\cite{nguyen2022flame}              & $43.22$ & $34.56$ & $58.24$ & $47.77$   \\
     & FLARE\cite{LiuP:C:2025c}                 & $64.92$ & $18.20$ & $84.25$ & $76.38$   \\
     & DEFEND\cite{LiuP:C:2025d}                & $\underline{69.60}^\dagger$ & $\underline{10.16}$ & $\underline{84.46}$ & $\underline{76.44}$   \\
     \rowcolor{teal!12}
     \cellcolor{white}& FedTrident (Ours)       & $\textbf{78.92}^\ast$ & $\textbf{2.92}$  & $\textbf{84.58}$ & $\textbf{76.56}$ \\
     \bottomrule
   \midrule

    \multirow{8}{*}{Material} 
     & FedAvg     & $60.16$ & $29.81$ & $77.73$ & $64.58$  \\
     & Krum       & $67.01$ & $25.28$ & $54.61$ & $45.08$  \\
     & TMean      & $56.93$ & $25.97$ & $72.91$ & $\underline{69.51}$  \\
     & Median     & $69.33$ & $23.66$ & $\underline{78.41}$ & $67.81$  \\   
     & FoolsGold  & $65.41$ & $18.61$ & $75.81$ & $66.29$ \\
     & FLAME      & $68.69$ & $13.59$ & $68.33$ & $60.17$ \\
     & FLARE      & $72.43$ & $13.33$ & $76.83$ & $67.39$  \\
     & DEFEND     & $\underline{75.07}$ & $\underline{9.04}$ & $77.51$ & $69.09$   \\
     \rowcolor{teal!12}
     \cellcolor{white}& FedTrident (Ours)      & $\textbf{84.16}$ & $\textbf{1.47}$ & $\textbf{78.80}$ & $\textbf{70.37}$  \\
    \bottomrule
   \midrule

    \multirow{8}{*}{Unevenness} 
     & FedAvg     & $38.83$ & $31.32$ & $65.40$ & $60.19$  \\
     & Krum       & $20.73$ & $49.33$ & $52.60$ & $46.49$  \\
     & TMean      & $42.67$ & $25.47$ & $67.64$ & $63.16$  \\
     & Median     & $43.98$ & $24.62$ & $67.62$ & $60.63$  \\   
     & FoolsGold  & $46.75$ & $22.47$ & $69.70$ & $62.92$ \\
     & FLAME      & $61.88$ & $20.96$ & $56.53$ & $47.13$ \\
     & FLARE      & $52.13$ & $22.70$ & $68.38$ & $63.37$  \\
     & DEFEND     & $\underline{63.40}$ & $\underline{10.57}$ & $\underline{70.91}$ & $\underline{64.35}$   \\
     \rowcolor{teal!12}
     \cellcolor{white}& FedTrident (Ours)   & $\textbf{81.70}$ & $\textbf{4.73}$ & $\textbf{72.82}$ & $\textbf{65.44}$ \\
    \bottomrule
  
  \end{tabular}
\begin{tablenotes} 
	\footnotesize
	\item$^\ast$ \textbf{Bold numbers are the best performance.}
	\item$^\dagger$ \underline{Numbers with underline are the second-best values.} 
\end{tablenotes}
\end{table}

The normalized confusion matrices of FedTrident for three tasks are illustrated in Fig.~\ref{fig_confusion_resnet} (with ResNet-18), Fig.~\ref{fig_confusion_efficientnet} (with EfficientNet-B1), and Fig.~\ref{fig_confusion_deit} (with DeiT-Tiny); source classes are marked in green and target classes are marked in red for readability. These confusion matrices exhibit strong diagonals across all classes, including the source–target pairs that define the TLFA objectives in each RCC task (water\(\rightarrow\)dry, gravel\(\rightarrow\)asphalt, and severe\text{-}uneven\(\rightarrow\)smooth). These results indicate consistently high per-class prediction accuracy and demonstrate clearly that \textit{FedTrident effectively prevents adversaries from realizing their targeted misclassifications that would otherwise jeopardize transportation safety}. Moreover, the nine matrices (three RCC tasks \(\times\) three DNN models) show uniformly strong performance, highlighting \textit{FedTrident’s stability and compatibility across tasks and models}.

\textbf{Observations on Model Type:} Lightweight models such as MobileNet-V3 and DeiT-Tiny show higher ASR with FedAvg and baseline defenses, indicating greater susceptibility to TLFAs. Their compact parameterization and limited representational redundancy make them more sensitive to localized neuron perturbations introduced by malicious clients. Despite the diversity of model types, ranging from convolutional (e.g., ResNet-18 and EfficientNet-B1) to transformer-based (DeiT-Tiny), FedTrident consistently achieves the best or second-best results across all metrics and tasks. Its SRE values remain high and ASR values remain low for most configurations, significantly outperforming baseline defenses. This demonstrates that \textit{FedTrident generalizes well across different feature extraction paradigms, not being tied to any specific model type}.

\begin{table*}[tbp]
  \scriptsize
  \centering
  \caption{The ablation results of FedTrident within default configurations. All values are ratios in \%.}
  \label{tab_results_ablation}
  \renewcommand{\arraystretch}{1.2} 
  \begin{tabular}{cc|cccc|cccc|cccc}
    \toprule
    \multicolumn{1}{c}{\multirow{2.5}{*}{Model Type}} &
    \multicolumn{1}{c|}{\multirow{2.5}{*}{Method}} &
    \multicolumn{4}{c|}{RCC @ Friction} &
    \multicolumn{4}{c|}{RCC @ Material} &
    \multicolumn{4}{c}{RCC @ Unevenness} \\
    \cmidrule(r){3-14}
    &   & SRE$\uparrow$& ASR$\downarrow$ & GAC$\uparrow$ & GAS$\uparrow$  & SRE$\uparrow$& ASR$\downarrow$ & GAC$\uparrow$ & GAS$\uparrow$  & SRE$\uparrow$& ASR$\downarrow$ & GAC$\uparrow$ &GAS $\uparrow$\\
    \midrule
    \multirow{3}{*}{ResNet-18} 
    & Detection  & $61.04$ & $14.04$ & $83.31$ & $72.42$    & $72.11$ & $10.24$ & $77.40$ & $68.78$    & $63.17$ & $11.22$ & $73.63$ & $66.48$ \\
     & Detection + Exclusion$^{\ddagger}$     & $\underline{75.02}^\dagger$ & $\underline{3.80}$ & $\underline{84.69}$ &\underline{74.39}  & $\underline{81.33}$& $\underline{4.53}$ & $\underline{79.54}$ &\underline{71.03}  & $\underline{79.85}$ & $\underline{8.80}$ & $\underline{73.71}$ &\underline{66.56} \\
     \rowcolor{teal!12}
     \cellcolor{white}& FedTrident (Full)     & $\textbf{88.04}^\ast$ & $\textbf{1.72}$ & $\textbf{85.16}$ & $\textbf{75.12}$  & $\textbf{87.47}$& $\textbf{1.89}$ & $\textbf{79.77}$ & $\textbf{71.76}$  & $\textbf{80.17}$ & $\textbf{5.68}$ & $\textbf{74.16}$ & $\textbf{67.21}$ \\
     \bottomrule
   \midrule

    \multirow{3}{*}{EfficientNet-B1} 
    & Detection  & $64.52$ & $11.04$ & $83.82$ & $73.41$    & $75.81$& $8.59$ & $\underline{78.99}$ & $70.10$   & $66.43$ & $11.93$ & $71.74$  & $64.32$ \\
     & Detection + Exclusion       & $\underline{81.60}$ & $\underline{5.32}$ & $\underline{84.09}$  & $\underline{74.17}$ & $\underline{82.99}$& $\underline{4.18}$ & $78.22$  & $\underline{71.52}$ & $\underline{75.78}$ & $\underline{8.90}$ & $\underline{72.26}$  & $\underline{64.71}$ \\
     \rowcolor{teal!12}
     \cellcolor{white}& FedTrident (Full)       & $\textbf{84.40}$ & $\textbf{3.92}$ & $\textbf{84.76}$ & $\textbf{75.38}$  & $\textbf{84.16}$& $\textbf{3.15}$ & $\textbf{79.50}$ & $\textbf{71.06}$  & $\textbf{80.75}$ & $\textbf{5.52}$ & $\textbf{72.91}$ & $\textbf{64.74}$ \\
    \bottomrule

   \midrule

    \multirow{3}{*}{DeiT-Tiny} 
    & Detection & $74.04$ & $9.20$ & $84.22$  & $72.45$  & $72.45$& $8.24$ & $74.19$  & $67.77$  & $67.18$ & $9.28$ & $69.98$  & $62.47$ \\
     & Detection + Exclusion       & $\underline{80.76}$ & $\underline{4.12}$ & $\underline{84.60}$  & $\underline{75.75}$ & $\underline{82.21}$& $\underline{4.77}$ & $\textbf{78.67}$  & $\underline{68.00}$ & $\underline{78.52}$ & $\underline{7.63}$ & $\underline{71.46}$  & $\underline{64.31}$ \\
     \rowcolor{teal!12}
     \cellcolor{white}& FedTrident (Full) & $\textbf{87.84}$ & $\textbf{1.00}$ & $\textbf{85.54}$ & $\textbf{78.12}$  & $\textbf{85.81}$& $\textbf{2.48}$ & $\underline{78.23}$ & $\textbf{69.60}$  & $\textbf{82.42}$ & $\textbf{2.33}$ & $\textbf{72.69}$ & $\textbf{65.33}$ \\
    \bottomrule  
  \end{tabular}
\begin{tablenotes} 
	\footnotesize
	\item$^\ast$ \textbf{Bold numbers are the best performance.}
	\item$^\dagger$ \underline{Numbers with underline are the second-best values.} 
    \item$^{\ddagger}$ \textit{Detection} denotes poisoned local model detection, and \textit{Exclusion} represents malicious vehicular client exclusion.
\end{tablenotes}
\end{table*}

\subsection{Resilience Analysis of FedTrident}\label{sec_resil_fedtrident}

\subsubsection{\textbf{Results on Different Malicious Client Rates}} 
As shown in Fig.~\ref{fig_poi_resnet} (with ResNet-18), Fig.~\ref{fig_poi_efficientnet} (with EfficientNet-B1), and Fig.~\ref{fig_poi_deit} (with DeiT-Tiny), increasing the proportion of malicious clients from 0\% to 40\% in steps of 10\% does not induce a commensurate rise in ASR when FedTrident is implemented. Take DeiT-Tiny as an example, FedTrident's ASR values remain within 1.00\%-5.60\% on Friction, 0.93\%-5.68\% on Material, and 2.33\%-7.78\% on Unevenness, which are always better than baselines. In contrast, FedAvg’s ASR values degrade sharply as the fraction of malicious clients increases: on Friction, it rises from 8.20\% to 47.08\%, on Material from 10.21\% to 46.16\%, and on Unevenness from 7.02\% to 54.93\%. Even for one of the best baseline defenses, FLARE, the increasing trend is worrying (7.01$\rightarrow$35.24\%, 8.95\%$\rightarrow$38.53\%, and 6.80\%$\rightarrow$34.57\%). These results underscore the strong sensitivity of TLFAs to attacker prevalence under both vanilla FL-RCC and state-of-the-art countermeasure, whereas \textit{FedTrident maintains low ASR values even at high malicious client rates, consistently across RCC tasks and model types}.

\subsubsection{\textbf{Results on Different Data Heterogeneity Levels}} 
As shown in Fig.~\ref{fig_noniid_efficientnet} (in the Friction task using EfficientNet-B1), we vary the IID level from 1.0 (strongly non-IID) to $+\infty$ (fully IID), in which the value is the parameter $\alpha$ in the Dirichlet distribution. The results show that the ASR values of FedAvg gradually increase from 18.52\% to 32.96\% when the data becomes more heterogeneous; accordingly, the SRE values degrade from 60.32\% to 38.12\%. Although one of the state-of-the-art defense baselines, FLARE, can always (no matter the IID level) improve model performance over FedAvg, it is still much worse than FedAvg-NA. As for the other state-of-the-art defense baseline, DEFEND, has significant improvement compared to FLARE; however, it is still a little bit worse than FedAvg-NA in term of ASR. Surprisingly, FedTrident is even slightly better than FedAvg-NA in all data distribution cases, as its ASR and SRE values remain at $[3.36\%, 3.92\%]$ and $[78.88\%, 84.40\%]$, respectively. Such stability and high performance clearly show that \textit{FedTrident is robust to varying levels of data heterogeneity}.  

\subsubsection{\textbf{Results on Multi-task}} 
Beyond separate RCC tasks, the performance of each method in many more multiple task classification that simultaneously involves friction, material, and unevenness, could be analyzed based on TABLE~\ref{tab_results_27}. The average SRE and ASR values of the best baseline are 66.22\% and 10.17\%, respectively; compared to FedAvg-NA (75.28\% and 0.28\%), the performance gap is 9.08\% and 9.89\% for ASR and SRE, respectively. As for FedTrident, it can sustain high SRE (74.39\%) and low ASR (0.61\%) values in such a challenging task. Note that the calculation of ASR and SRE relies on the source and target classes; if the total number of classes is huge (e.g., 27 in our case), it is difficult for the FL-RCC system to identify and thwart TLFAs. \textit{The neuron-wise analysis of FedTrident could recognize the attack goal of TLFAs successfully, thus benefiting the FL-RCC system in such a complex task}.

\subsubsection{\textbf{Results on Dynamic Attacks}} 
Instead of static attacks whose source and target classes are fixed during FL training, we consider dynamic attacks where adversaries can change the two classes based on their attack goal. The setting in the Friction task is to flip from \textit{water} to \textit{wet} in the first 10 rounds, then from \textit{water} to \textit{dry} in the last 50 rounds. Likewise, the interim classes are gravel and slight-uneven in Material and Unevenness, respectively. As summarized in TABLE~\ref{tab_results_dynamic}, all defense baselines struggle to deal with the dynamic attacks, and their worst performance could be 3.12\%, 67.36\%, 52.60\%, and 45.08\% for SRE, ASR, GAC, and GAS, respectively. FedTrident can always achieve the best, leading to average improvements (compared to the best baseline) of 12.24\% for SRE, 6.88\% for ASR, 0.81\% for GAC, and 0.83\% for GAS. \textit{These results demonstrate that even under dynamic attacks, the detection module of FedTrident can accurately identify the source and target classes in each round; thus, the exclusion and remediation modules can also function effectively}.   

\subsubsection{\textbf{Ablation Study}} 
The poisoned local model detection serves as the basic module of FedTrident that the second module (malicious vehicular client exclusion) relies on; the final module, corrupted global model remediation, takes the exclusion module's results as input. To evaluate the effectiveness of each module, we present an ablation study of FedTrident in Table~\ref {tab_results_ablation}. Results show that combining the detection module with the exclusion module can, on average, improve performance by 11.26\%, 4.64\%, 1.11\%, and 1.36\% for SRE, ASR, GAC, and GAS, respectively, compared to only using the detection module. Moreover, the full FedTrident, integrating all three modules, achieves the overall best performance; compared to detection + exclusion, SRE, ASR, GAC, and GAS values are further optimized by 4.78\%, 2.71\%, 0.61\%, and 0.88\%, respectively. Such a comparison shows that \textit{all three modules in FedTrident can help improve performance}.  


\section{Conclusions}\label{sec_con}
We propose FedTrident for FL-RCC against TLFAs, with three modules: 1) poisoned local model detection based on neuron-wise analysis to identify target goals and GMM to distinguish poisoned models from benign ones; 2) malicious vehicular client exclusion adjusting client ratings based on model-level detection to preclude persistent poisoning; and 3) corrupted global model remediation that capitalizes on machine unlearning to correct the already-corrupted global model after each exclusion. Holistic evaluation involving various RCC tasks and DNN models indicates that under TLFAs, FedTrident is superior to state-of-the-art countermeasures, and could achieve the same performance as an attack-free scenario. Moreover, FedTrident is resilient to diverse malicious client rates, varying data heterogeneity levels, complex tasks, and dynamic attacks.

We will extend FedTrident to safeguard other ITS tasks beyond RCC, e.g., world model-based driving. Moreover, we will investigate TLFAs in the context of self-supervised FL and adopt FedTrident to such situations. Finally, we will have more practical considerations, e.g., executing TLFAs and evaluating FedTrident in real-world large-scale vehicular communication and computation environments. 
 
\section*{Acknowledgments}
This work was supported in parts by WASP, VR, and in kind by the KAW Foundation granting access to Berzelius at the National Supercomputer Centre.


%

\bibliographystyle{IEEEtran}
\bibliography{TDSC-RCC}


\vspace{5pt}

\begin{IEEEbiography}[{\includegraphics[width=1in,height=1.25in,clip,keepaspectratio]{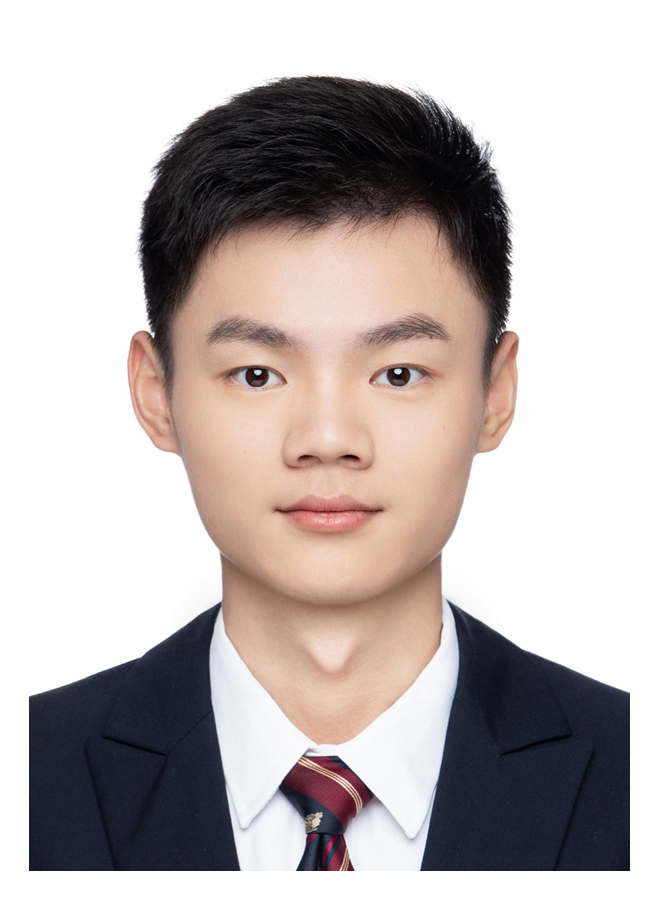}}]
{Sheng Liu} (Graduate Student Member, IEEE) received the B.Eng. degree and the
master’s degree from the School of Intelligent Systems Engineering, Sun Yat-sen University,
China, in 2021 and 2024, respectively. He is currently pursuing a Ph.D. degree with the Networked Systems Security (NSS) Group at KTH Royal Institute of Technology, Stockholm, Sweden. His research interests include trustworthy AI, federated learning, security, privacy, and intelligent transportation.
\end{IEEEbiography}

\vspace{5pt}

\begin{IEEEbiography}[{\includegraphics[width=1in,height=1.25in,clip,keepaspectratio]{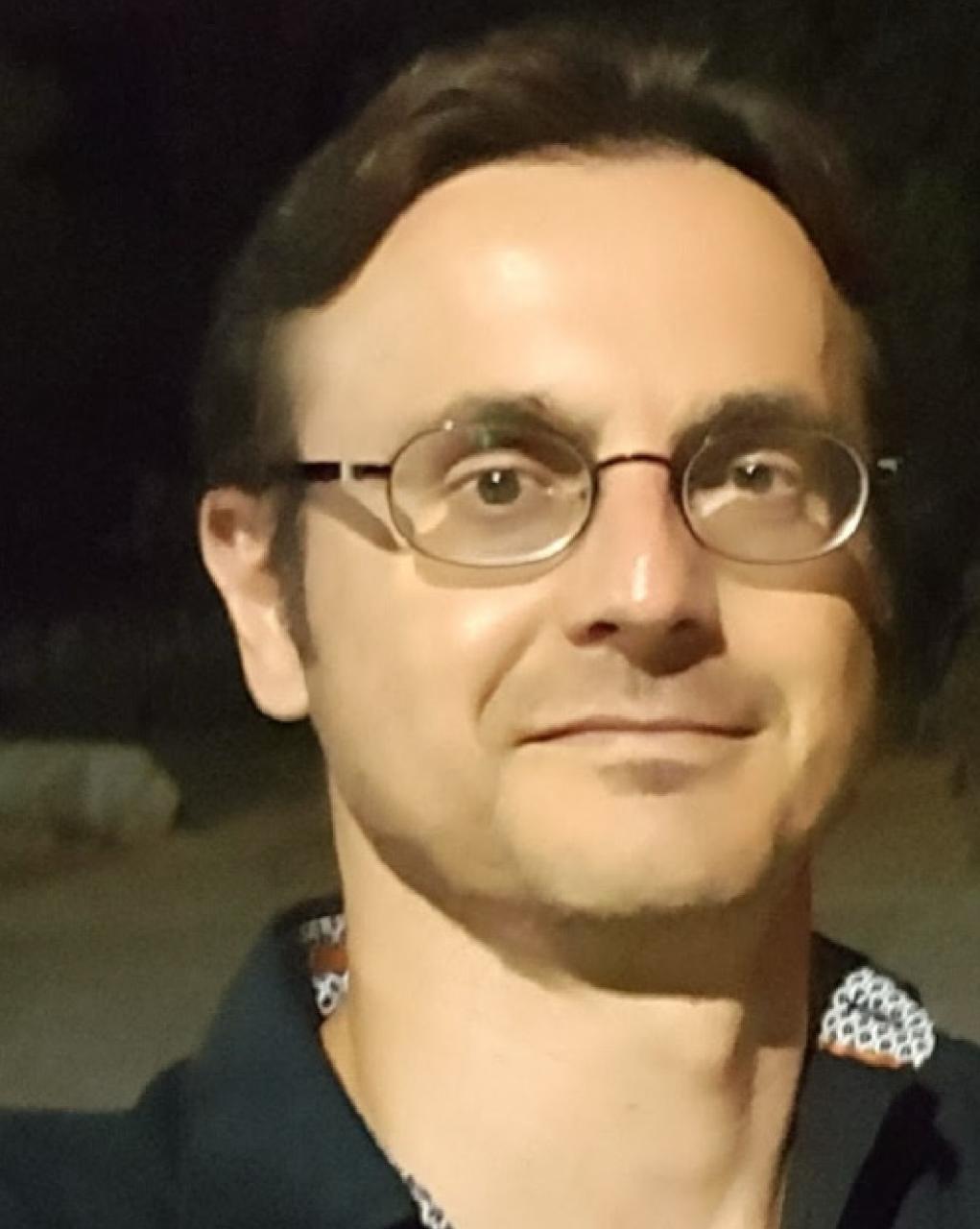}}]
{Panagiotis Papadimitratos} (Fellow, IEEE) received the Ph.D. degree from Cornell University, Ithaca, NY, USA. At KTH Royal Institute of Technology, Stockholm, Sweden, he leads the Networked Systems Security (NSS) Group and is a member with the Steering Committee of the Security Link Center. He serves or served as a member for the ACM WiSec and CANS conference steering committees and the PETS Editorial and Advisory Boards; the Program Chair for ACM WiSec’16, TRUST’16, and CANS’18 conferences; the General Chair for ACM WISec’18, PETS’19, and IEEE EuroS\&P’19 conferences; an Associate Editor for IEEE TRANSACTIONS ON MOBILE COMPUTING, IEEE/ACM TRANSACTIONS ON NETWORKING, and IET Information Security journals; and the Chair for the Caspar Bowden PET Award. He is a fellow of the Young Academy of Europe, a Knut and Alice Wallenberg Academy Fellow, and an ACM Distinguished Member.
\end{IEEEbiography}

\vfill

\end{document}